\documentclass[a4paper,fleqn,usenatbib,useAMS]{mnras}
\usepackage[final]{graphicx}
\usepackage{amssymb}
\usepackage{amsmath}
\usepackage{lastpage}
\usepackage{dsfont}
\usepackage{float}
\usepackage{caption}

\usepackage{subcaption}
\usepackage{rotating}

\usepackage{tabularx}
\usepackage{pdfpages}
\usepackage{color}
\usepackage[toc,page]{appendix}
\usepackage{gensymb}
\usepackage{epstopdf, epsfig}
\usepackage{hyperref}
\usepackage{pgfplots}
\pgfplotsset{compat=newest}
\usepackage{multicol} 
\usepackage{bm}	
\usepackage{pdflscape}
\usepackage[T1]{fontenc}
\usepackage{ae,aecompl}
\usepackage{times}

\title{Resilient habitability of nearby exoplanet systems}
\date{\today}
\author[G. Kokaia, M. B. Davies, A. J. Mustill]{Giorgi Kokaia$^{1}$\thanks{Contact e-mail: \href{mailto:giorgi@astro.lu.se}{giorgi@astro.lu.se}}, Melvyn B. Davies$^{1}$, Alexander J. Mustill$^{1}$
\\
$^{1}$Lund Observatory, Department of Astronomy and Theoretical Physics, Lund University, Box 43, SE-221 00 Lund, Sweden}
\date{\today}

\pubyear{2019}

\begin{document}
\label{firstpage}
\pagerange{\pageref{firstpage}--\pageref{lastpage}}
\maketitle

\begin{abstract}
We investigate the possibility of finding Earth-like planets in the habitable zone of 
34 nearby FGK-dwarfs, each known to host one giant planet exterior to their habitable zone 
detected by RV. First we simulate the dynamics of the planetary systems in their 
present day configurations and determine the fraction of stable planetary orbits within 
their habitable zones. Then, we postulate that the eccentricity of the giant planet is a result of an instability in their past during which one or more other planets were ejected from the system. We simulate these scenarios and investigate whether planets orbiting in the habitable zone survive the instability.
Explicitly we determine the fraction of test particles, originally found in the 
habitable zone, which remain in the habitable zone today. We label this fraction the 
{\it resilient habitability} of a system. We find that for most systems the 
probability of planets existing [or surviving] on stable orbits in the habitable zone 
becomes significantly smaller when we include a phase of instability in their history. 
We present a list of candidate systems with high resilient habitability for future observations. 
These are: HD~95872, HD~154345, HD~102843, HD~25015, GJ~328, HD~6718 and HD~150706. The known planets in the last two systems have large observational uncertainties on their eccentricities, which propagate into large uncertainties on their resilient habitability. Further observational constraints of these two eccentricities will allow us to better constrain the survivability of Earth-like planets in 
these systems.
\end{abstract}


\section{Introduction}\label{sec:intro}

One of the key goals for exoplanet science is to observe an Earth-like planet located
in the habitable zone of a Sun-like star, where
the habitable zone is defined as the spherical shell at a distance that permits liquid water to exist on the planet's surface. 

Detecting Earth-mass (i.e. low-mass) planets has been technologically limited, however new instruments such as ESPRESSO (currently operational with the sensitivity slowly increasing~\citealp{Pepe2010, Leite2018}), space missions like PLATO~\citep{Rauer2014} and future large telescopes such as E-ELT and TMT~\citep{Gilmozzi2007, Skidmore2015} offer the prospect for the discovery of Earth-like planets in the near future.

Discovering Earth-like planets will, however, be difficult even with the next generation of instruments. 
One must focus attention on  a relatively small number of  stars (or planetary systems). 
A natural question is then: {\it which systems do we select for further detailed study?}
In this paper we will consider local stars possessing one observed (massive) planet
located beyond the habitable zone. Through numerical modelling of their dynamical past,
we select the subset having the best prospects for possessing a habitable low-mass planet.
Considering systems containing a Jupiter-mass planet further out makes sense for a number of reasons:

\begin{enumerate}
\item Recent work has shown that the presence of Jupiter was instrumental in shaping the 
inner parts of the Solar System by clearing out or preventing the delivery of a large 
fraction of the solids in the early inner disc. This might be a requirement for systems 
such as the Solar System to form~\citep{Batygin2015, Childs2019, Lambrechts2019}, rather 
than the systems commonly found by \textit{Kepler} which consist of super-Earths and 
mini-Neptunes on tightly packed orbits within 0.5 au~\citep{Mullally2015}.

\item Until relatively recently there was commonly held belief that Jupiter protected 
the Earth from impacts (see e.g.~\citealp{Ward2000}). This belief has come under scrutiny 
as the effect of Jupiter on impact rates has been looked at more closely through numerical 
simulations~\citep{Laakso2006}. Jupiter actually appears to cause a slight enhancement in 
impacts from bodies originating in the asteroid and Kuiper belt~\citep{Horner2008, Horner2009} 
whilst reducing the impact rate of long-period comets~\citep{Horner2010}. Regardless, giant 
planets appear to have been instrumental in delivering water and other volatiles to the inner 
parts of the system, whether it happened after formation
\citep{Chyba1990, Morbidelli2000, Abramov2009, CarterBond2014, Grazier2016} or during 
formation \citep{Raymond2006, Raymond2017, OBrien2018}.

\item Systems with gas giants are very likely to also host low-mass planets. Recent studies 
have shown that $\sim 40$\% of systems that have low-mass planets (in this case ``low-mass'' 
implies $\lesssim10\mathrm{M}_\oplus$) also host an external gas giant 
\citep{Zhu2018, Bryan2019}. Further, \cite{Zhu2018} invert this occurrence rate using 
overall occurrence rates for each type of planet and conclude that $\sim90$\% of gas 
giant systems should have interior, lower-mass planets. The correlations found are not 
strictly for the Earth-like planets that we consider in this work. However, one could 
argue that if there were data to consider these planets then the correlation might be 
even stronger, given the fact that a gas giant might be required to even be able to have 
Earth-like planets. For comparison, Burke et al. (2015) determined the occurrence 
rate for Earth-like planets around GK-dwarfs given by extrapolation from 
\emph{Kepler} systems to be $\sim6\pm3$\% for planets with radii between $0.75$ and $1.75$ 
Earth radii and 300--700 day orbits. By instead considering \emph{Kepler} non-detections,
 Hsu et al. (2019) determined an upper bound ($84^\mathrm{th}$ percentile) 
of 27\% on the occurrence rate of planets with 
radii between $0.75$ and $1.5$ Earth radii on periods of 237--500 days. 
They caution however that the bound is likely overestimated by up to a factor of two due to 
contamination by false alarms.

\item The RV-signal strength from a planet is proportional to 
$M_{\rm p}\sin I_{\rm p}$ where $I_{\rm p}$ is the inclination of the orbital 
plane with respect to the plane of the sky, with the observer at $I_{\rm p}=90^\circ$. 
For nearby giant planets, $Gaia$ will be able to constrain the orbital plane 
\citep{Sozzetti2008, Perryman2014,Ranalli+18}. Given that planetary systems 
seem to typically have low mutual 
inclinations \citep{Lissauer2011, Johansen2012}, we can further constrain which systems 
to observe by picking out those in which the giant planet has $I_{\rm p}\approx90^\circ$. 
This gives a strong prior for the amplitude of the RV signal from the Earth-like planet 
being observed to be close to its maximum possible value.
\end{enumerate}

However, a gas giant may destabilise certain orbits including those  within the habitable zone of a
planetary system~\citep{Jones2001, Menou2003, VerasArmitage05, 
Matsumura2013, Agnew2017, Agnew2018, Georgakarakos2018}. In this paper we will not only 
re-examine the present day habitability of systems 
containing giant planets, but 
also determine the fraction of test particles remaining within  habitable zones when one accounts
for the past dynamical evolution of particular systems. We term this fraction the
\emph{resilient habitability} \citep{Carrera2016} of these systems

The resilient habitability of a system is determined by assuming that the system initially consisted 
of multiple planets and that the eccentricity of the observed planet is the result of one or more 
other planets' being ejected when the system became unstable. The idea is that changes to the 
orbits and ejection of planets during the instability will do considerably more damage to 
Earth-like planets in the habitable zone than the system as
seen today with its post-instability architecture. 
In summary, from our simulations we determine two types of habitability for each of the observed
systems we study:
\begin{itemize}
    \item[] \textbf{Present-day habitability:} The fraction of test particles that remain in the habitable zone when the system is simulated in its present day configuration. We denote this as $f_{\rm hab,1P}$.
    \item[] \textbf{Resilient habitability:}  The fraction of test particles that remain in the habitable zone when the past dynamical evolution of the system is simulated. We denote this as $f_{\rm rhab}$.
\end{itemize}

This paper is arranged as follows.
In section~\ref{sec:gp} we discuss how scattering between two planets can lead to the ejection of one
with the other becoming more bound and eccentric. We show how the eccentricity reached is a function
of the mass ratio of the two planets.
In section~\ref{sec:sel} we present the known planet-hosting systems we have selected to investigate, each of which has a single giant planet exterior to the habitable zone. Section~\ref{sec:num} describes 
our calculations of both the present day and resilient habitability of these systems. 
The results of these calculations are presented in
section~\ref{sec:result}. The implications of our results are discussed in section~\ref{sec:disc}.
In section~\ref{sec:obs}  we list the systems which in our view represent the best candidates for further searches
for Earth-like planets. We summarise the paper in section~\ref{sec:su}.

\section{Giant planet scattering}\label{sec:gp}

We know from observations \citep{Buchhave2018}, and suspect from simulations 
\citep{Thommes2008, Mordasini2015}, that giant planets rarely form alone. Furthermore, it has 
been shown that giant planets form on circular or nearly circular orbits 
\citep{Bitsch2013, Ragusa2018}  which are later made more eccentric, whereas the observed 
exoplanet eccentricity distribution is consistent with most of them having been made eccentric 
through planet--planet scattering \citep{Ford2008, Juric2008}.

In this section we consider scattering between massive planets within two-planet systems. 
When two planets are placed on sufficiently close, low-eccentricity orbits, they will become
unstable, leading to the two orbits crossing which then leads to a series of strong planet--planet
scattering events, with one planet being ejected whilst one remains on a more bound, and eccentric,
orbit.  For a review on the dynamical evolution of planetary systems see~\cite{Davies2014}.

\begin{figure}
\centering
\includegraphics[width = 1\columnwidth]{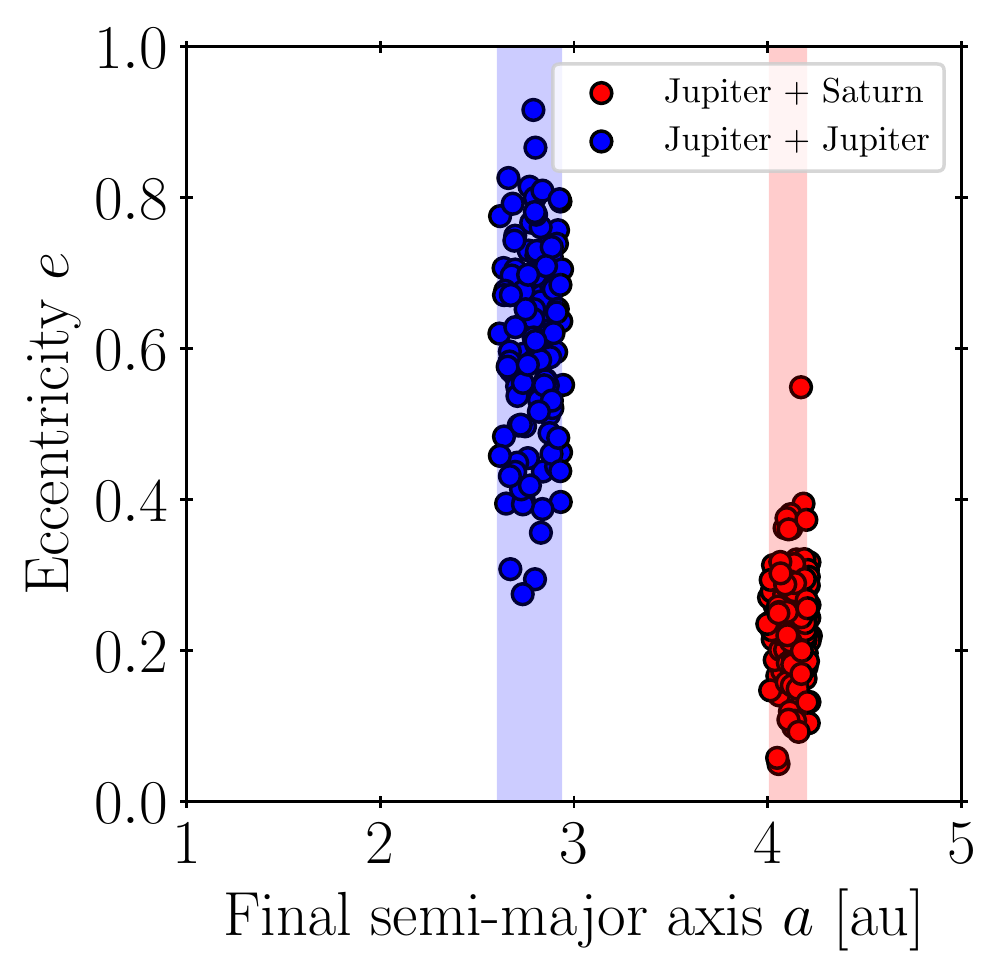}
\caption{The figure shows the $a$ and $e$ of a surviving Jupiter-mass planet  after having ejected a 
  second planet in an unstable two-planet system. The shaded region shows the range 
  of semi-major axes permitted based on a simple energy consideration. The 
two colours show two different mass combinations: Jupiter plus Jupiter, and Jupiter plus Saturn.}
\label{fig:scatters}
\end{figure}

Here,  we model two systems orbiting around a Solar-mass star: one containing 
two Jupiter-mass planets and one containing one Jupiter-mass
planet and one Saturn-mass planet, in order to investigate how the eccentricity distribution of the 
planets left behind (once one planet is ejected) depends on the mass ratio of the two planets.
We place one Jupiter at 5.2 au and place a second either Jupiter- or Saturn-mass planet in the range [5.2,7.6] au. This setup eventually leads to planet--planet scattering during which either of the Jupiter-mass planets can be ejected in the first case and the Saturn-mass planet is ejected in the second case.
Figure~\ref{fig:scatters} shows the semi-major axis and eccentricity of the remaining planet for the two systems.
The distribution of eccentricities is clearly a function of the mass ratio of the two planets.
In ejecting a second Jupiter, the remaining Jupiter-mass planet has to give up more energy and angular momentum, thus making it more bound and more eccentric 
than when the inner Jupiter ejects a Saturn-mass planet.

We will make use of the mass-ratio dependence of eccentricity when we consider the dynamical
histories of our selected observed systems. For example, if the observed planet in a system
has a eccentricity below about 0.3, we can then expect the planet to have ejected (only) Saturn-mass
planets in the past, whereas for larger eccentricities, the planet is more likely to have ejected 
one (or more) planets of equal mass.

\section{System selection}\label{sec:sel}

We select systems from the exoplanet archive\footnote{https://exoplanetarchive.ipac.caltech.edu/} 
with a single gas giant ($0.3\mathrm{M}_\mathrm{J}<M_{\rm p}\sin I_{\rm p}<8\mathrm{M}_\mathrm{J}$ 
where $\mathrm{M}_\mathrm{J}$ is the mass of Jupiter) orbiting an FGK-dwarf star 
($1.4\mathrm{M}_\odot\gtrsim \mathrm{M}_\star \gtrsim0.65\mathrm{M}_\odot$, $\mathrm{R}_\star<2R_\odot$) beyond its habitable zone (HZ). 

The habitable zone (HZ) is defined as a shell at a distance from a star with a given luminosity 
($L$) and effective temperature ($T_\mathrm{eff}$) in which a planet can have liquid water on 
its surface. The boundaries of the HZ for each system are determined using the formula derived by~\cite{Kopparapu2014}, shown below:
\begin{equation}
d = \left(\frac{L/L_\odot}{S_{\rm eff}}\right)^{0.5}~\rm AU
\label{eq:hz}
\end{equation}
\noindent where $L_\odot$ is the Solar luminosity and
\begin{equation}
S_{\rm eff}=S_{\rm eff\odot}+aT_\star+bT^2_\star+cT^3_\star+dT^4_\star
\label{eq:seff}
\end{equation}
is a measure of the flux received by the planet and 
$T_\star=T_\mathrm{eff}-T_\odot$ ($T_\odot=5780~K$). The values for the parameters 
are given in table~\ref{tab:params}. The inner edge of the HZ is set using the model 
parameters for runaway greenhouse habitability, the limit at which the temperature of 
the planet, under any realistic atmospheric conditions, inevitably becomes too
high. The outer edge is set using the maximum greenhouse parameters, which is the 
limit at which an optimally constructed atmosphere to maximise the greenhouse effect 
can no longer prevent all the water from freezing.

\begin{table}
\centering
\caption{Parameters used in equation~\ref{eq:seff} for determining the boundaries of the habitable zone as given by \protect\cite{Kopparapu2014}}
\begin{tabular}{l r r }
 &\multicolumn{1}{c}{Inner edge}&\multicolumn{1}{c}{Outer edge}\\
Parameter & Runaway Greenhouse & Maximum Greenhouse\\
\hline
$S_{\rm eff\odot}$ & $1.107$ & $0.356$\\
$a$ & $1.332\times10^{-4}$ & $6.171\times10^{-5}$\\
$b$ & $1.58\times10^{-8}$ & $1.698\times10^{-9}$\\
$c$ & $-8.308\times10^{-12}$ & $-3.198\times10^{-12}$\\
$d$ & $-1.931\times10^{-15}$ & $-5.575\times10^{-16}$
\label{tab:params}
\end{tabular}
\end{table}

To further limit the number of systems we simulate, we impose another criterion; the pericentre calculated using an eccentricity one standard deviation lower than the mean observed value must lie exterior to the HZ.

\begin{figure}
\centering
\includegraphics[width = 1\columnwidth]{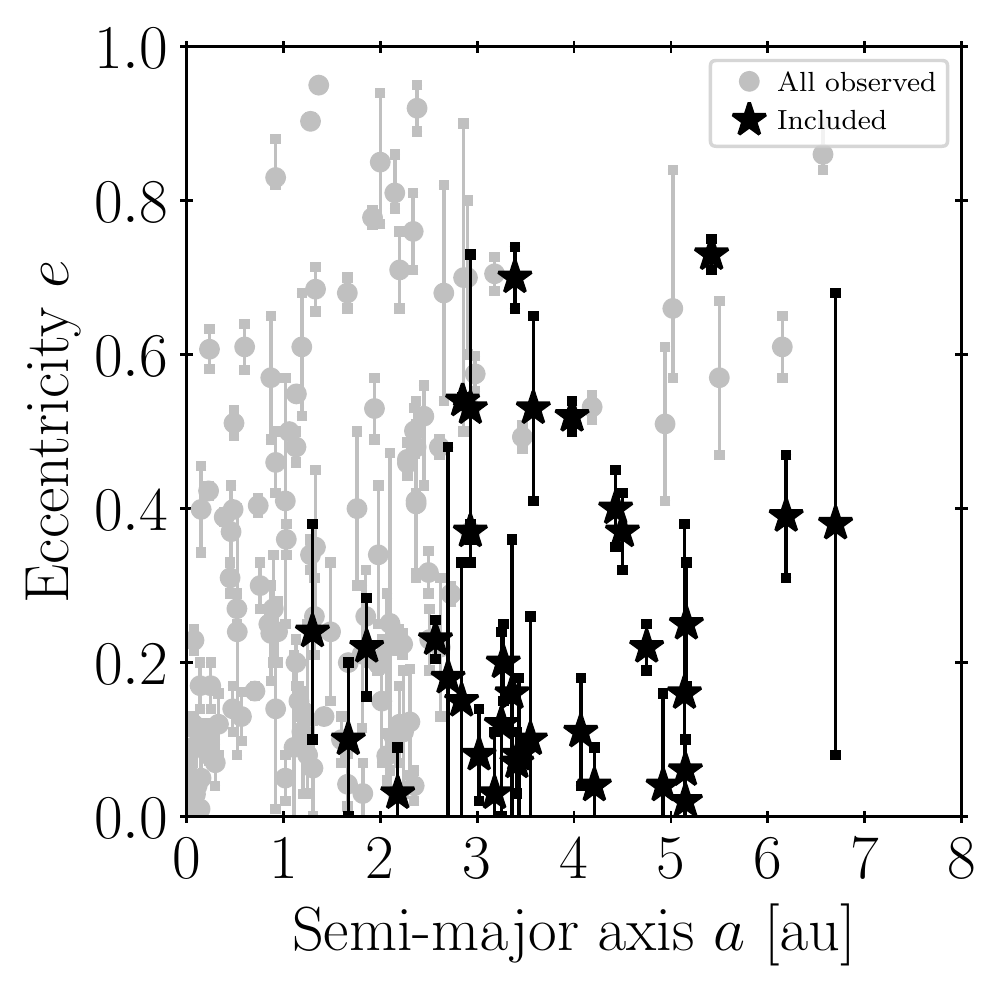}
\caption{All of the known single-planet RV systems that meet our selection criteria for 
  planet mass, and for stellar mass and radius. 
  The black stars in the plot show the systems in table~\ref{tab:sysparam}; 
  the grey markers are systems which meet the selection criteria 
  but where the planet has an orbit that enters, is inside or interior to the habitable zone.}
\label{fig:singles}
\end{figure}

The criteria leave us with 34 systems as listed in table~\ref{tab:sysparam}. We include in this table
the stellar parameters used to determine the HZ. They are plotted and highlighted in figure~\ref{fig:singles} along with every other known single planet RV system that matches our criterion for planetary mass and stellar type.
Half of these systems can be found on the short-list of~\cite{Agnew2017, Agnew2018} who investigated the present day stability of orbits in the HZ. For the remaining half, some planets were considered  by~\cite{Agnew2017, Agnew2018}  to be  so distant that the orbits of test particles in the HZ would be unaffected by the planet, and therefore were not investigated. We also include some high eccentricity systems with limited stability that did not make their cut and some systems discovered subsequent to their work.

\begin{table*}
\begin{center}

\caption{The stellar parameters used to determine the habitable zone using the runaway greenhouse and maximum greenhouse parameters from \protect\cite{Kopparapu2014} and the planetary orbits. The effective temperatures, parallaxes and luminosities are all from Gaia DR2 \protect\citep{GaiaDR2}; the planetary parameters and stellar mass are determined in the references below. The table is ordered by the distance to the midpoint of the habitable zone.
}
\begin{tabular}{l r r r r r r r r}
 & \multicolumn{1}{l}{Stellar} &$T_{\rm eff}$ &Luminosity & \multicolumn{1}{l}{Stellar} &HZ mid- & Planet Mass & Semi-major & \\
System & Mass ($\mathrm{M}_\odot$) & (K) & $(\mathrm{L}_\odot)$ & Parallax (mas) & point (AU) & ($\mathrm{M}_\mathrm{J}\sin I$) & axis (AU) & Eccentricity\\
\hline
${}^1$HD 114613 & 1.36 & 5709 & 4.44 & 49.3 & 2.79 & 0.48 & 5.16 & $0.25\pm0.08$\\
${}^2$HD 222155 & 1.21 & 5720 & 3.22 & 19.7 & 2.37 & 2.12 & 5.14 & $0.16\pm0.22$\\
${}^2$HD 72659 & 1.43 & 5898 & 2.38 & 19.2 & 2.01 & 3.85 & 4.75 & $0.22\pm0.03$\\
${}^{3} \psi^1$ Dra B & 1.19 & 6213 & 2.11 & 43.9 & 1.85 & 1.53 & 4.43 & $0.4\pm0.05 $\\
${}^4$HD 25171 & 1.09 & 6127 & 2.00 & 18.0 & 1.81 & 0.95 & 3.02 & $0.08\pm0.06$\\
${}^2$HD 24040 & 1.11 & 5805 & 1.89 & 21.4 & 1.80 & 3.86 & 4.92 & $0.04\pm0.12$\\
${}^2$HD 13931 & 1.30 & 5837 & 1.74 & 21.1 & 1.72 & 2.20 & 5.15 & $0.02\pm0.04$\\
${}^5$HD 220689 & 1.04 & 5944 & 1.50 & 21.3 & 1.59 & 1.06 & 3.36 & $0.16\pm0.20$\\
${}^2$HD 89307 & 1.27 & 5955 & 1.36 & 31.2 & 1.51 & 2.11 & 3.27 & $0.2\pm0.05$\\
${}^2$HD 86226 & 1.06 & 5924 & 1.23 & 21.9 & 1.44 & 0.92 & 2.84 & $0.15\pm0.18$\\
${}^{6}$HD 68402 & 1.12 & 5950 & 1.20 & 12.7 & 1.42 & 0.81 & 2.18 & $0.03\pm0.06$\\
${}^2$HD 27631 & 0.94 & 5732 & 1.10 & 19.9 & 1.38 & 1.45 & 3.25 & $0.12\pm0.12$\\
${}^7$HD 150706 & 1.17 & 5920 & 1.07 & 35.3 & 1.35 & 2.71 & 6.70 & $0.38\pm0.3$\\
${}^{6}$HD 152079 & 1.10 & 5722 & 1.04 & 11.3 & 1.34 & 2.18 & 3.98 & $0.52\pm0.02$\\
${}^2$HD 70642 & 0.96 & 5667 & 1.02 & 34.1 & 1.34 & 1.85 & 3.18 & $0.03\pm0.08$\\
${}^8$HD 32963 & 0.94 & 5746 & 1.01 & 26.2 & 1.32 & 0.70 & 3.41 & $0.07\pm0.04$\\
${}^5$HD 6718 & 1.08 & 5730 & 0.96 & 18.2 & 1.29 & 1.68 & 3.55 & $0.10\pm0.08$\\
${}^5$HD 142022 & 0.90 & 5487 & 0.89 & 29.1 & 1.27 & 4.44 & 2.93 & $0.53\pm0.20$\\
${}^{9}$HD 171238 & 0.81 & 5440 & 0.96 & 22.3 & 1.22 & 2.72 & 2.57 & $0.23\pm0.03$\\
${}^{2}$14 Her & 0.90 & 5282 & 0.71 & 55.7 & 1.15 & 4.66 & 2.93 & $0.37\pm0.01$\\
${}^{10}$HD 102843 & 0.72 & 5346 & 0.72 & 15.9 & 1.15 & 0.36 & 4.07 & $0.11\pm0.07$\\
${}^4$HD 30669 & 0.92 & 5386 & 0.71 & 17.5 & 1.14 & 0.47 & 2.70 & $0.18\pm0.30$\\
${}^2$HD 290327 & 0.84 & 5545 & 0.73 & 17.7 & 1.14 & 2.43 & 3.43 & $0.08\pm0.10$\\
${}^{9}$HD 98736 & 0.92 & 5312 & 0.65 & 30.8 & 1.10 & 2.33 & 1.86 & $0.22\pm0.064$\\
${}^2$HD 95872 & 0.70 & 5180 & 0.57 & 132.2 & 1.04 & 3.74 & 5.15 & $0.06\pm0.04$\\
${}^2$HD 154345 & 0.71 & 5468 & 0.60 & 54.7 & 1.04 & 3.07 & 4.21 & $0.04\pm0.05$\\
${}^{11}$HD 42012 & 0.83 & 5342 & 0.54 & 27.1 & 1.00 & 1.60 & 1.67 & $0.07\pm0.07^\dagger$\\
${}^{12}$HD 25015 & 0.86 & 5079 & 0.41 & 26.7 & 0.89 & 4.48 & 6.19 & $0.39\pm0.08$\\
${}^{13} \epsilon$ Eri & 0.83 & 4975 & 0.38 & 312.3 & 0.86 & 1.55 & 3.39 & $0.7\pm0.04$\\
${}^5$HD 166724 & 0.81 & 5101 & 0.39 & 22.1 & 0.86 & 3.53 & 5.42 & $0.73\pm0.04$\\
${}^2$HD 87883 & 0.67 & 4964 & 0.34 & 54.6 & 0.81 & 1.54 & 3.58 & $0.53\pm0.12$\\
${}^{14}$HD 164604 & 0.80 & 4684 & 0.26 & 25.4 & 0.73 & 2.7 & 1.3 & $0.24\pm0.14$ \\
${}^{2}$BD-17 63 & 0.72 & 4669 & 0.22 & 29.0 & 0.67 & 2.85 & 2.85 & $0.54\pm0.01$\\
${}^{15}$GJ 328 & 0.69 & 3989 & 0.10 & 48.7 & 0.48 & 2.30 & 4.50 & $0.37\pm0.05$\\

\label{tab:sysparam}

\end{tabular}
\end{center}

\begin{flushleft}
${}^1$\protect\cite{Wittenmyer2014}, ${}^2$\protect\cite{Stassun2017}, ${}^{3}$\protect\cite{Endl2016}, ${}^4$\protect\cite{Moutou2011}, ${}^5$\protect\cite{Marmier2013}, ${}^{6}$\protect\cite{Jenkins2017},${}^7$\protect\cite{Boisse2012}, ${}^8$\protect\cite{Rowan2016}, ${}^{9}$\protect\cite{Ment2018}, ${}^{10}$\protect\cite{Feng2019} ${}^{11}$\protect\cite{Rey2017}, ${}^{12}$\protect\cite{Rickman2019}, ${}^{13}$\protect\cite{Benedict2006} ${}^{14}$\protect\cite{Robertson2013}, \\
${}^{\dagger}$Note: HD 42012 is given with a $3\sigma$ upper bound of 0.2. We reformulate it as in the table to make consistent with the rest of the systems.
\end{flushleft}
\end{table*}

\section{Setup of the numerical experiments}\label{sec:num}

We perform all our \textit{N}-body simulations using the Bulirsch--Stoer integrator of the \textsc{Mercury} package \citep{Chambers1999}, with an accuracy parameter of $10^{-12}$ resulting in a relative energy error of less than $10^{-5}$ for all runs.\\

\subsection{Present-day habitability}

As a first step we simulate all the planetary systems in their present day configuration to determine the present day habitability, i.e. the fraction of test particle orbits which keep their semi-major axis in the HZ despite the presence of a nearby giant planet. We refer to this value as $f_{\rm hab,1P}$.

We place 100 test particles in the HZ with their semi-major axes distributed uniformly from inner to outer edge. We place them on circular orbits and make them co-planar with the planet. We randomly pick the three Keplerian angles (argument of perihelion, $\omega$; longitude of ascending node, $\Omega$, and mean anomaly, $M$) between $0$ and $360^\circ$ and then integrate the systems for $10^7$ years. For each system we check if 10 Myr was sufficient for it to relax, i.e. if after 10 Myr particles are still being lost. If any test particle was lost within the last Myr we run the system for another Myr and check again repeating the process until no more test particles are lost. We find that for most systems 10 Myr is sufficient. For each system we do three simulations. One simulation using the reported eccentricity of the giant planets and two using the $\pm 1 \sigma$ eccentricity-values shown in table~\ref{tab:sysparam}.

\begin{figure*}
  \begin{center}
  
  \begin{subfigure}[t]{0.33\textwidth}
    \centering
    \includegraphics[width=\linewidth]{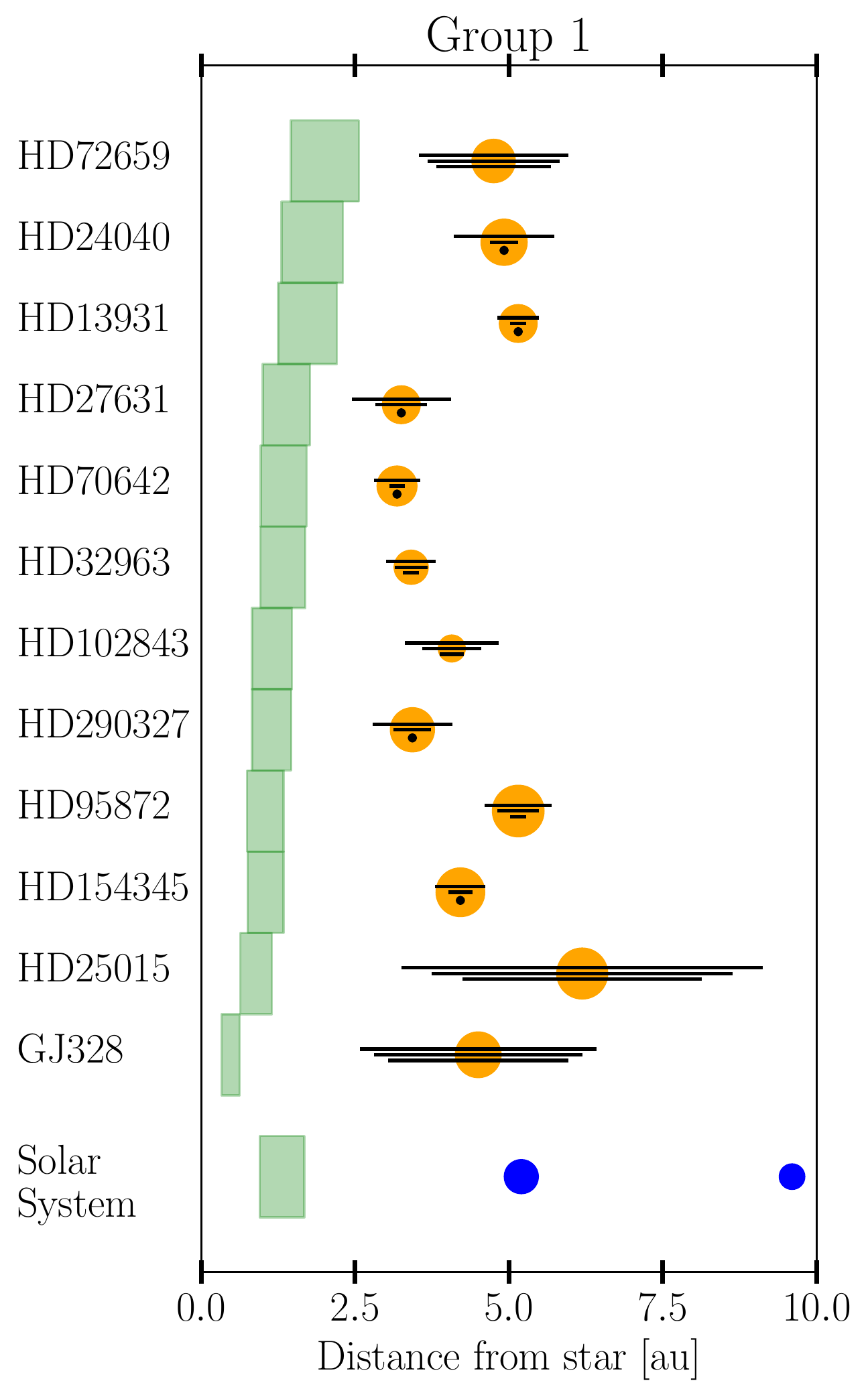} 
  \end{subfigure}
  \begin{subfigure}[t]{0.33\textwidth}
    \centering
    \includegraphics[width=\linewidth]{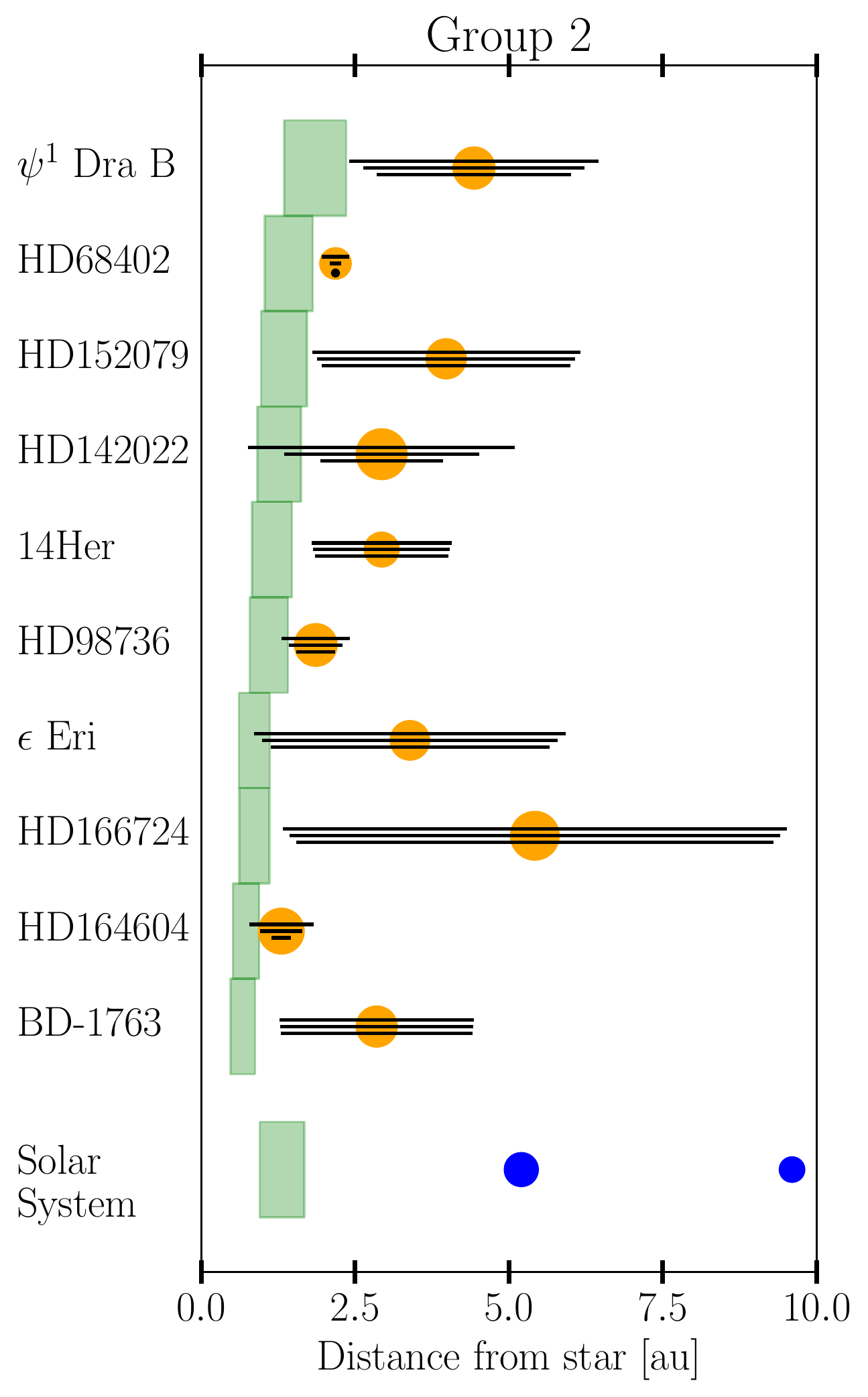} 
  \end{subfigure}
  \begin{subfigure}[t]{0.33\textwidth}
    \centering
    \includegraphics[width=\linewidth]{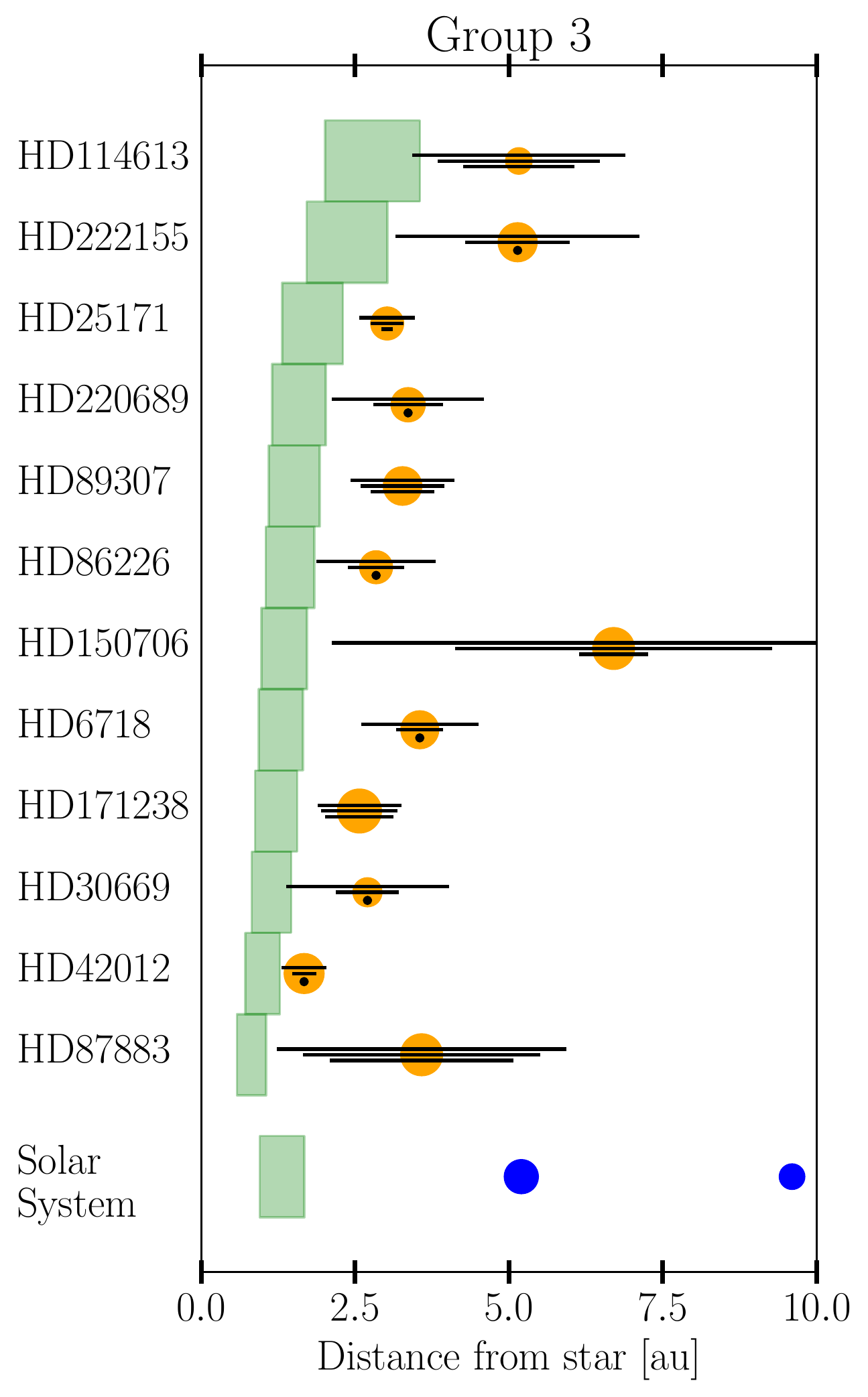} 
  \end{subfigure}
  \end{center}

  \caption{ The planetary systems studied in our simulations. 
    The planet is shown as an orange circle, where the area of 
    the circle  is proportional to $\sqrt{M_{\rm p}/M_\star}$. 
    The black lines show the peri- and apo-centre range of the orbit using 
    the reported and $\pm~1\sigma$ values of the eccentricity.
    If the lower limit is consistent with zero, this is instead 
    represented by a black dot.
    The green shaded area is the habitable zone calculated according to the 
    formula from \protect\cite{Kopparapu2014} (Equations~\ref{eq:hz} and~\ref{eq:seff} 
    of this paper). The systems are split into three categories, from left to right: 
    Group 1 comprises systems with planets on distant orbits and/or with low 
    eccentricity; Group 2 comprises systems with planets on 
    nearby orbits and/or with high eccentricity; 
    and Group 3 comprises systems where the planetary orbits have 
    uncertain eccentricities that could place
    the systems in either Group 1 or Group 2. The Solar System is shown 
    for scale at the bottom of each panel.}
  \label{fig:current_systems}
\end{figure*}

The systems we simulate are shown in figure~\ref{fig:current_systems}. They are split into three groups:
\begin{itemize}
\item[]Group 1: Systems where the planet has a considerably larger semi-major axis than 
  the outer edge of the HZ and a low eccentricity, or has a moderate eccentricity with 
  a much larger semi-major axis.
\item[]Group 2: Systems where the planet has a semi-major axis just outside the outer edge of the HZ with a low to moderate eccentricity or at larger semi-major axis with a sufficiently high eccentricity to bring the pericentre very close to the HZ.
\item[]Group 3: Systems that could fit in either Group 1 or 2, due to the planets' having large uncertainties in their eccentricities.
\end{itemize}

\subsection{Resilient habitability}\label{sec:reshab}

\subsubsection{Planets}

To determine the the resilient habitability we need to perform simulations that can represent the past dynamical evolution of an observed system. We postulate that any simulation of the system's past that ends with a
massive planet having an eccentricity consistent with observations can be used to represent the system's history.

Figure~\ref{fig:scatters} shows that the outcomes of planet--planet scattering between planets of two different mass ratios (1:1 and 3:10) covers the entire range of observed eccentricities. We take these two mass ratios and construct two sets of initial conditions that contain planets with those ratios plus additional ones in order to capture secular effects during and after scattering. We call them 3E and 4H and they are set up as follows:
\begin{itemize}
    \item[] \textbf{3E}: This set contains three equal-mass planets, all with a mass equal to the $M_{\rm p}
     \sin I_{\rm p}$ of the observed planet.
    \item[] \textbf{4H}: This set contains four planets with hierarchical masses, following the Solar System giants, with the innermost planet having a mass equal to the $M_{\rm p} \sin I_{\rm p}$ of the observed planet. We will refer to planets in this set as [lower case] jupiter, saturn, uranus, neptune.
\end{itemize}
We randomly pick the three Keplerian angles uniformly between $0$ and $360^\circ$ and draw random initial eccentricities uniformly from 0 to 0.01. The initial inclinations are drawn randomly between 0 and 5$^\circ$ which results in mutual inclinations between 0 and 10$^\circ$ with an average around 2-3$^\circ$: 
this is consistent with observations of Kepler multiple systems~\citep{Lissauer2011, Johansen2012}. The planet radius is set following \cite{Bashi2017}. Planets below $0.39\mathrm{M}_\mathrm{J}$ follow a relation of $R\propto M^{0.55}$ and planets with masses above it follow $R\propto M^{0.01}$ with the breakpoint radius being $1.1\mathrm{R_J}$.

For each set in each system we place the innermost planet at two times the observed semi-major axis and initial separation between the planets is set to 3 mutual Hill radii in the 3E runs and 3.5 mutual hill radii for the 4H runs. 
The mutual Hill radius of two planets with masses $M_1$ and $M_2$ orbiting a star with mass $M_\star$ at semi-major axes of $a_1$ and $a_2$ is defined as:
\begin{equation}
R_{\rm MHill} = \left(\frac{M_1+M_2}{3M_\star}\right)^{1/3}\frac{a_1+a_2}{2}.
\label{eq:hill}
\end{equation}

This small spacing is chosen in order to trigger the instability on a short time-scale, and to be computationally efficient. The time until the instability is triggered does not affect the duration of the scattering phase, as that is set by the orbital time-scale of the planets~\citep{Malmberg2011}. Nor does it affect broadly the properties
of the systems once they have stabilised following the ejection of one or more planets.

\subsubsection{Test particles}

In two-planet scattering one can narrowly constrain the expected range of final semi-major axes of the surviving planet by considering energy conservation (as seen in figure~\ref{fig:scatters}). When more planets are added, the final energy of the planets becomes less predictable and the expected range of semi-major axes is widened. Given that we are modelling a system where the planet has a specific semi-major axis we need to circumvent this. We do this by utilizing the fact that planetary dynamics are essentially scale free; i.e. if one takes a system and doubles all the semi-major axes it would almost always have the same exact dynamical outcome. The one exception is the probability of collisions, which decreases with increasing semi-major axis \citep{Ford2008}. We find them to be rare in our simulations ($<3\%$) so this is not a concern. Given this, we can rescale the semi-major axes of the remaining planets and the location of HZ according to the semi-major axis of the planet in the system we are modelling.

The post facto rescaling means that a larger range of test particles has to be included in the simulation. We place approximately 450 test particles (circular, coplanar and randomly phased) between 0.8 times the inner edge of the HZ to 2.5 times the outer edge. The exact number is determined so the spacing between them is the same as in the previous experiment. Any realization (run) that ends with the rescaled location of the HZ outside where initially there were test particles is discarded. 

Two changes are made to the simulation to ensure efficiency. Firstly, the ejection distance is set to 250 au which ensures that test particles which clearly are not going to be habitable are removed in a timely fashion. Second, we we remove any test particles that pass within 0.05 au of the host star.
This removes highly eccentric test particles that 
would significantly slow down the integration during their pericentre passages.

We then run the simulations for $10^7$ years. Our initial conditions trigger the instability in less than $10^5$ years in all systems and the scattering phase rarely lasts more than $10^6$ years. This means that $10^7$ years is usually more than enough for the system to evolve and the phase of scattering to end. However if there are test particles being ejected in the last Myr we run the simulation until there has been a Myr with no ejections.

The resilient habitability for each run is calculated by rescaling the location of the HZ by $a_{\rm final}/a_{\rm obs}$, checking which particles started with their semi-major axes in it and then seeing what fraction are left at the end.

%
%

\section{Results}\label{sec:result}

As explained earlier, we considered the dynamical evolution of two kinds of planetary systems:
3E (where we have three planets of the same mass) and 4H (where we have four planets
following the same ratio of masses of our own gas giants). For each system we 
investigated, we ran 100 simulations for each
architecture following a phase of instability and planetary scattering: this ejected some planets and
left the inner planet on a more eccentric orbit. In Section~\ref{sec:gp}, we 
showed that scattering of a Jupiter and a Jupiter, and of a Jupiter and a Saturn, 
can between them span the observed eccentricity range. Therefore, we first consider the question:
{\it what blend of 3E and 4H runs best matches the observed eccentricity distribution 
of all observed giant exoplanets?}

 We determine the weighting by comparing the eccentricity distribution we find for the innermost planets in our simulations with observations. Our observational sample is taken from the exoplanet archive\footnote{https://exoplanetarchive.ipac.caltech.edu/ as of February 11th 2019}. We select giant planets with masses in the range $[0.3, 8]\mathrm{M}_\mathrm{J}$ and semi-major axes $a>1.3$ au orbiting stars with masses $[0.65, 1.4]\mathrm{\,M}_\odot$ resulting in a sample of 177 planets.

\begin{figure}
\centering
\includegraphics[width = 1\columnwidth]{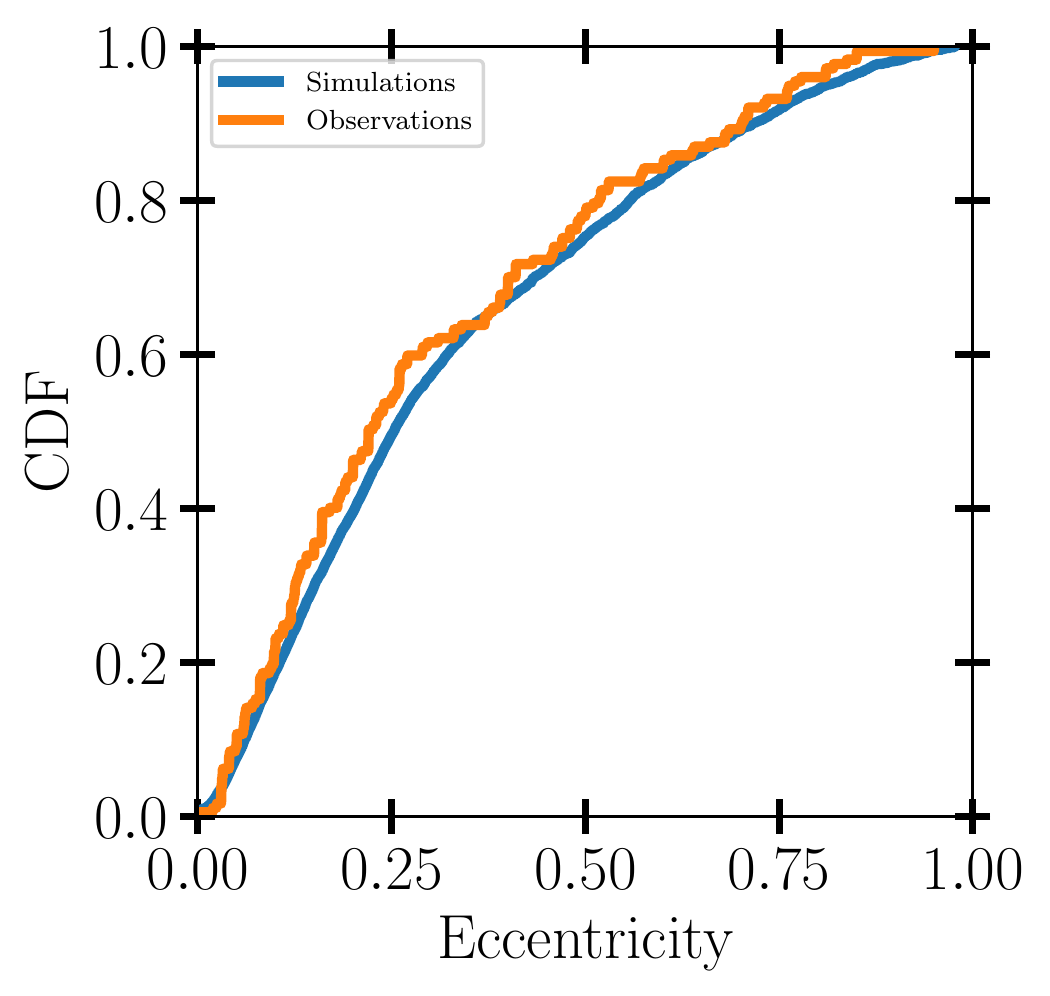}
\caption{The cumulative distribution of exoplanet eccentricities. The observed RV 
  sample is shown in orange. In blue we show the distribution 
  arising from a 1:1 blend of the results of our two sets of simulations, 3E and 4H.
  For the 3E simulations we use the eccentricity of the innermost planet, while for 
  the 4H simulations we use the eccentricity of jupiter which in 
  $\sim97\%$ of the cases is also the innermost planet.}
\label{fig:blends}
\end{figure}

\begin{figure}
\centering
\includegraphics[width = 1\columnwidth]{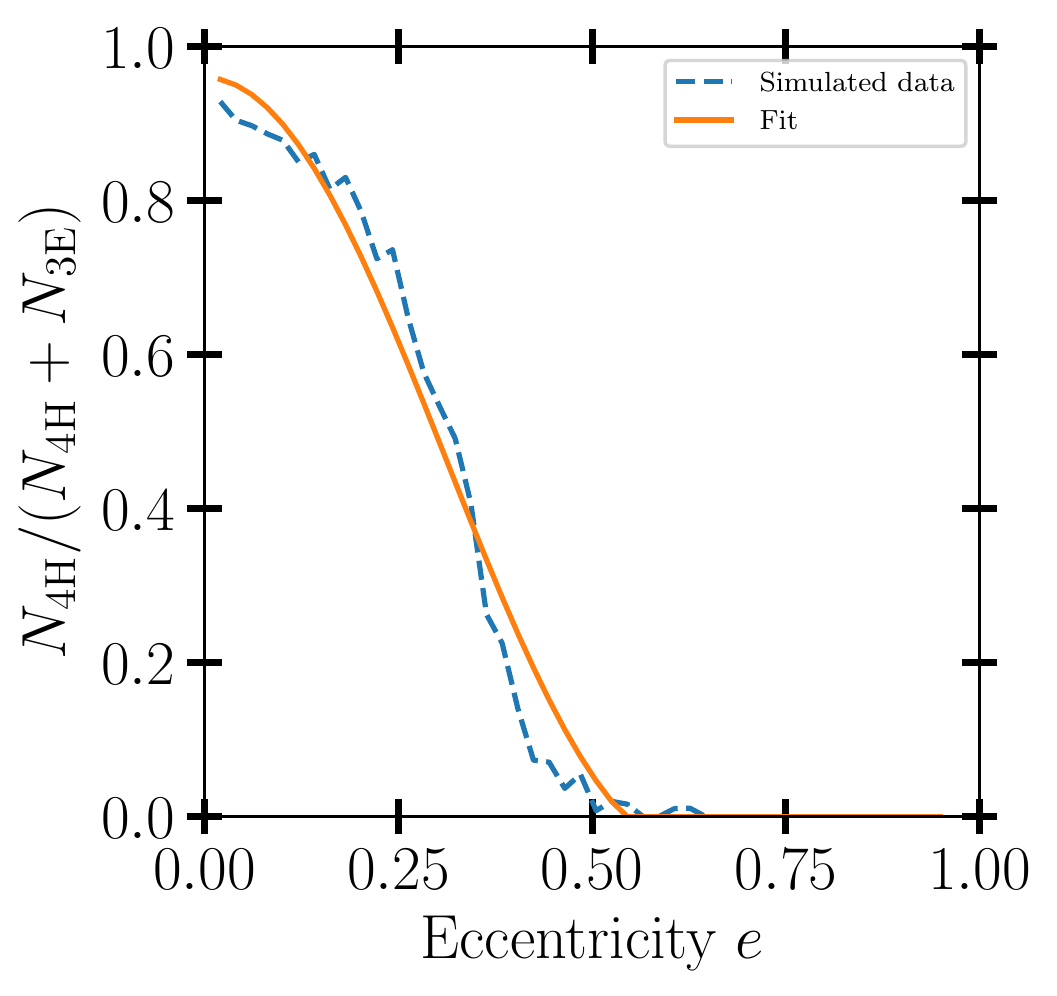}
\caption{Here we show the fraction of initially 4H systems as a function of final eccentricity, 
  when using the 1:1 global blend shown in Figure~\ref{fig:blends}. 
  The blue dashed line shows the results from the simulations, while the solid 
  orange line shows a polynomial fit.}
\label{fig:fracs}
\end{figure}

We generate 1000 eccentricity distributions for a range of 3E:4H ratios from the simulated eccentricities by randomly sampling them without replacement. We then perform a 2-sample KS-test between the observed eccentricities and our generated distribution. We find that randomly drawn distributions with 3E:4H ratios between 2:3 and 5:4 are consistent with the observations in more than 95\% of the cases, and we chose to use a 1:1 ratio. The comparison is shown in figure~\ref{fig:blends}, and the resulting fraction of initially 4H systems as a function of eccentricity is shown in figure~\ref{fig:fracs}. Figure~\ref{fig:fracs} can also be interpreted as the probability of a system with a given reported eccentricity to initially have been a 4H system.

For each of the systems illustrated in  figure~\ref{fig:current_systems}, we modelled dynamical
histories using both the 3E (three equal-mass planets) and 4H (four planets following the same
ratio of masses of our own gas giants). For each system, we performed 100 runs for each of the 3E 
and 4H architectures. For further analysis, keeping those where the 
inner planet's orbital eccentricity matched that of
the observed system (within the 1$\sigma$ uncertainties for eccentricity).

We used the distribution shown in figure~\ref{fig:fracs} to weight the 3E and 4H runs for each system to determine its $f_{\rm rhab}$-distribution. 
Often, the number of 3E and 4H runs that fall within the eccentricity range for a system do not match the expected fraction from 
figure 5, given its eccentricity. This is because when determining the 
expected fraction as a function of the reported eccentricity we use the eccentricities at the end of our simulations ($e_{\rm final}$), whereas we allow for eccentricity 
oscillations (i.e. angular momentum exchange via secular interactions) when determining the $f_{\rm rhab}$-distribution. We allow for the oscillations because a number of runs would have ended up in the correct eccentricity range if the simulation had been stopped at a different time. After taking this into account we up-sample 
either the included 3E or 4H runs to give the correct ratio for the reported eccentricity of the observed system.
 Once we know which runs to use and how to re-sample them we determine the $f_{\rm rhab}$-distribution.

\begin{figure*}
\centering
\includegraphics[width = 1\textwidth]{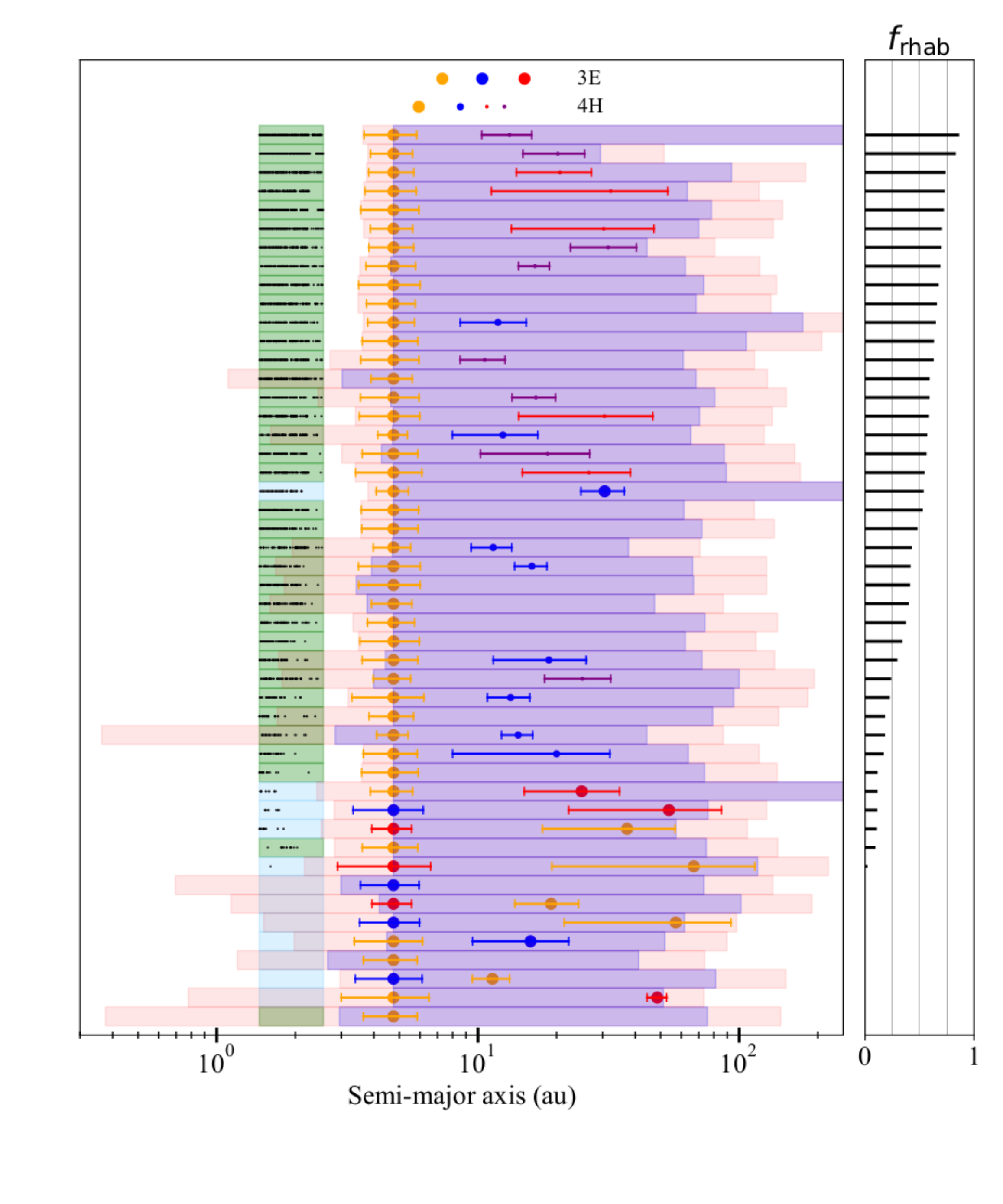}
\caption{The figure shows the outcome of a set of runs for HD~72659. The habitable 
  zone is shown by the shaded rectangle around 2\,au: green for the 4H 
  runs and cyan for the 3E runs. Coloured circles show the final semimajor 
  axes of the planets, with the error bars indicating their 
  pericentre and apocentre values. The planets' initial locations, 
  with the same colouration, are shown at the top of the figure.
  The final semimajor axes of the test particles are shown as small 
  points, blue for those that contribute to $f_\mathrm{rhab}$ and 
  grey for those that do not. The shaded blue region shows the 
  range of the planets' semimajor axes during the integration, with 
  output sampled every 10\,kyr, and the peach region shows the 
  range of the planets' pericentres and apocentres, also sampled 
  every 10\,kyr. Finally, the panel to the right shows the 
  $f_\mathrm{rhab}$ value for each system.}
\label{fig:fullpage}
\end{figure*}

\begin{figure*}
  \begin{center}
  
  \begin{subfigure}[t]{0.33\textwidth}
    \centering
    \includegraphics[width=\linewidth]{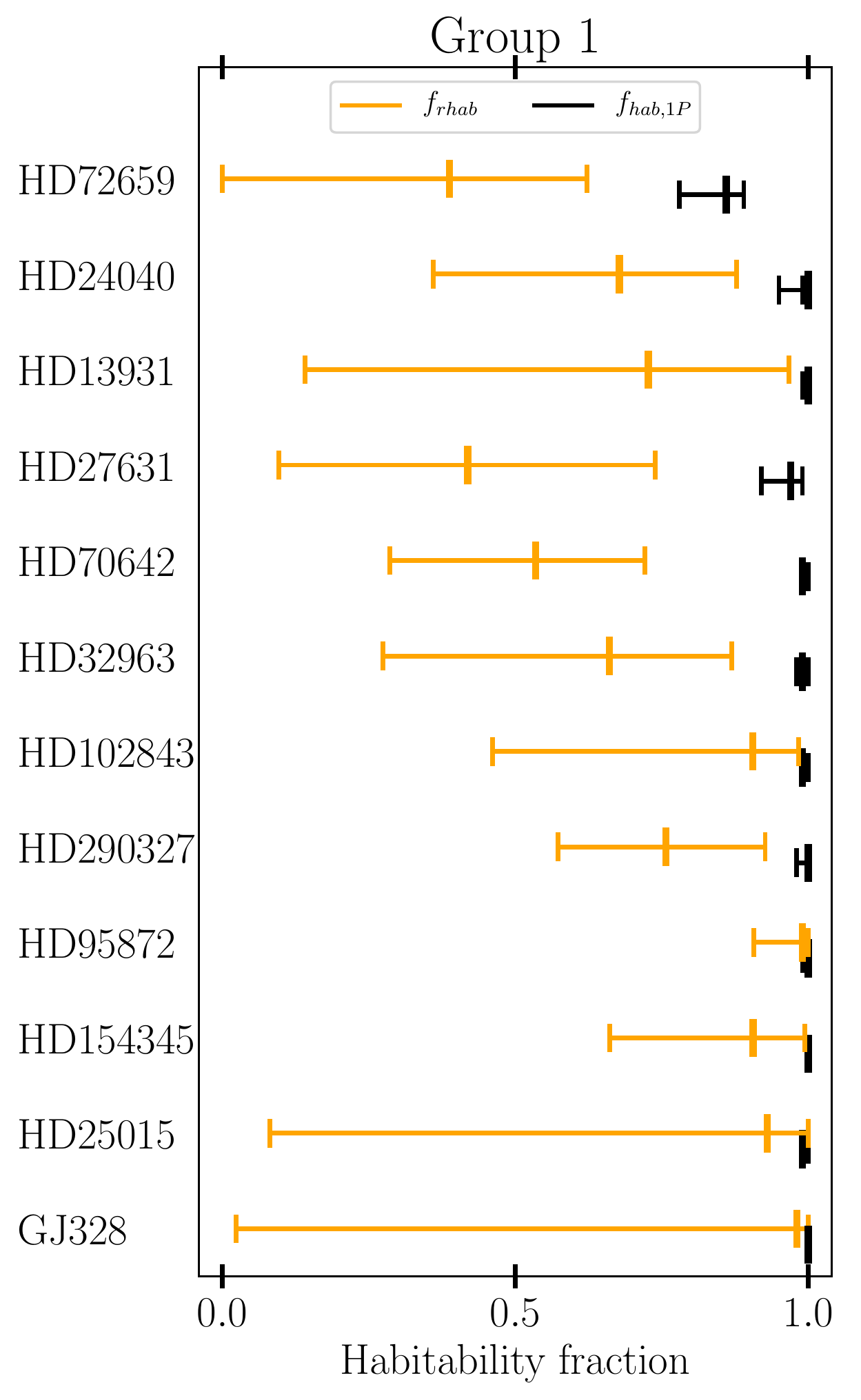} 
  \end{subfigure}
  \begin{subfigure}[t]{0.33\textwidth}
    \centering
    \includegraphics[width=\linewidth]{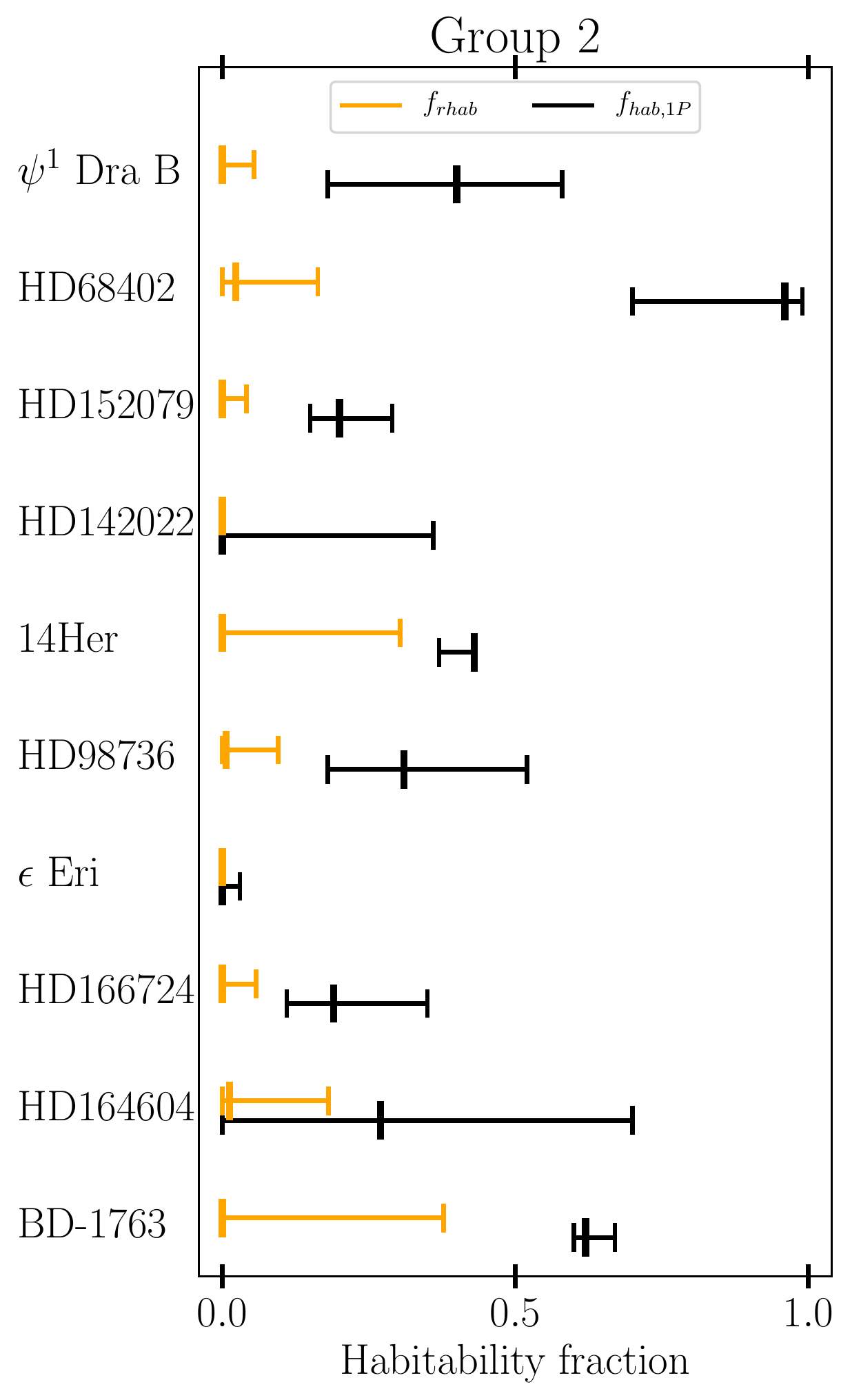} 
  \end{subfigure}
  \begin{subfigure}[t]{0.33\textwidth}
    \centering
    \includegraphics[width=\linewidth]{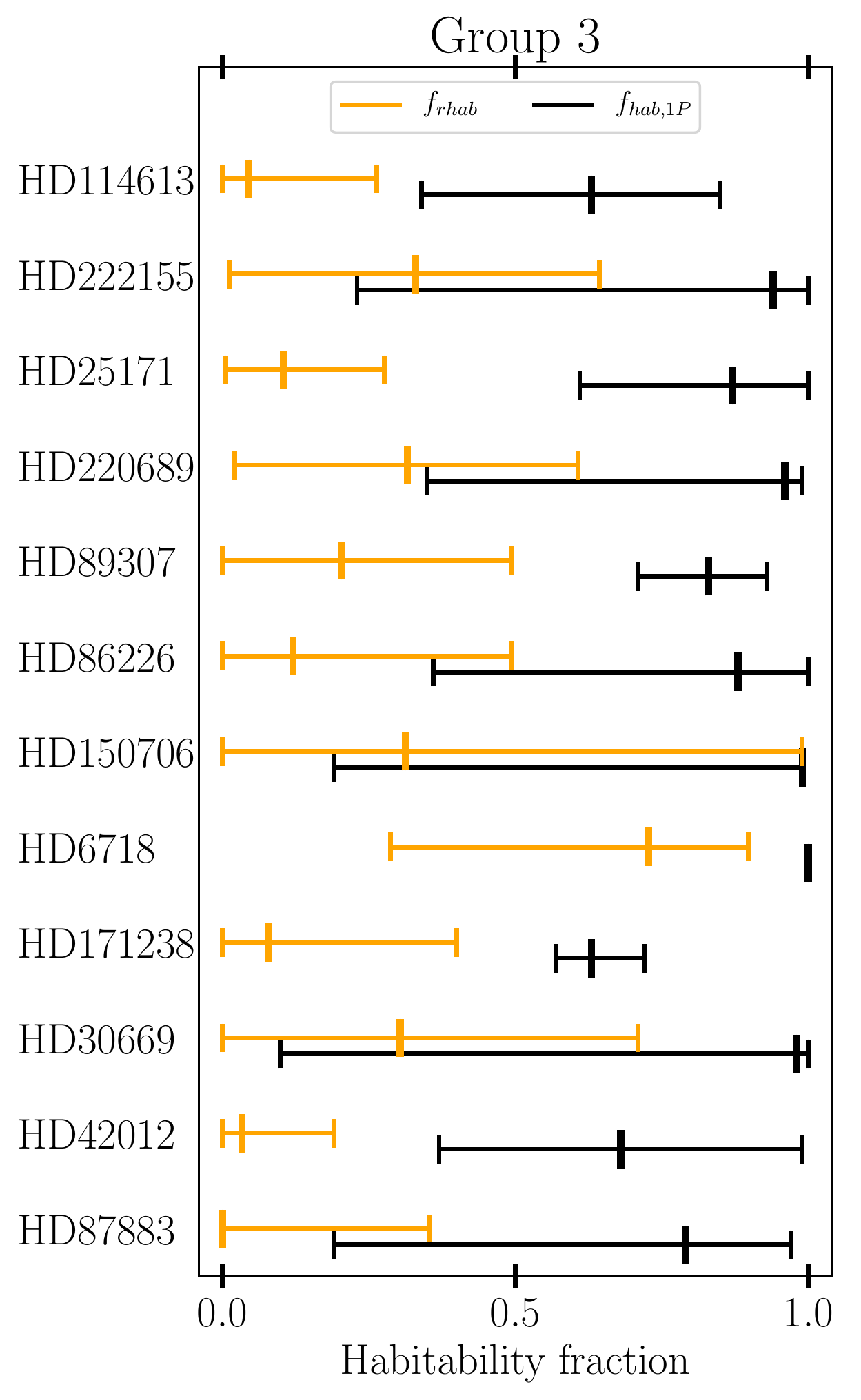} 
  \end{subfigure}
  \end{center}
	
\caption{The present day and resilient habitability of all the systems. 
  The black line indicates the present day habitability of the systems, in the 
  configuration shown in figure~\ref{fig:current_systems}. The midpoint on the line 
  show the survival fraction found when simulating using the reported 
  eccentricity and the endpoints are the result of using $\pm 1 \sigma$ values. 
  The orange line shows the median and ``$1~\sigma$ range'' (i.e., the 
  central 68\% of the distribution) 
  of the $f_{\rm rhab}$ distribution for each system, using the 
  blending ratio given by the reported eccentricity (see Figure~\ref{fig:fracs}). }
\label{fig:frhab}

\end{figure*}

Figure~\ref{fig:fullpage} shows the outcome of a subset of runs simulating HD 72659. 
The resilient habitability, i.e.\ the fraction of the test particles originally in the habitable
zone (HZ) remaining within the HZ at the end of the run, $f_{\rm rhab}$, 
for each run is shown in the panel on the right. 
The figure includes a correctly-weighted  combination of 4H and 3E runs, differentiated by the colour of the 
HZ where green represents 4H and light blue 3E. All of the runs have 
eccentricities consistent with the reported 
eccentricity, when we include secular eccentricity oscillations as described above. 
We show the configurations of the remaining planets for each of 48 runs, sorted in decreasing $f_{\rm rhab}$.
Additionally we show at the top the initial 
configurations for 3E and 4H. The colour labelling for the initial configuration and the final
48 systems is the same.
For each planet, the location of the  circle denotes the semi-major axis whilst the error
bars denote the maximum and minimum distances form the host star. 
The light backgrounds show the range of the giant planet semi-major axes (light blue) or positions (light red) over the course of each run, sampled every 10\,kyr.

We see a clear difference between the 3E and 4H runs: 4H gives a broad range of $f_{\rm rhab}$-values, even when looking at runs with similar eccentricities: while
the upper quartile have $f_{\rm rhab} \approx 0.75 -1 $, others show much smaller values of $f_{\rm rhab}$.
 On the other hand, for the 3E runs we see that in most cases $f_{\rm rhab}\approx0$; we see that the vast 
 majority of the twelve 3E runs shown are located at the bottom of the distribution. Massive planets have
 directly invaded the HZ in about one half of the 3E runs, leading in all these cases to the essentially
 complete removal of test particles. However, we see a few examples of 4H runs where the HZ is invaded
 without equally destructive outcomes. This is because the planet in question was of much lower
 mass (roughly that of Neptune).
 In a large fraction of  runs we can see that test particles have been ejected even though a planet never entered the HZ. This, along with the gaps seen in the test particle distribution in nearly every run, 
can be explained by resonances which are pumping up the eccentricities of the test particles. A detailed discussion of the resonances will follow in section~\ref{sec:resonaces}.

\subsection{$f_{\rm rhab}$ and $f_{\rm hab, 1p}$ distributions for the three groups}

As can be seen in figure~\ref{fig:fullpage}, runs for a given system can show a very broad range
of values of $f_{\rm rhab}$. Therefore for each system, we plot the range of values of $f_{\rm rhab}$
(the $1\sigma$-equivalent range, i.e., the central 68\%  of the distribution) 
in figure~\ref{fig:frhab},
together with the present day habitability simulations, $f_{\rm hab,1P}$, in black. 
The systems are separated into the three Groups as shown in figure~\ref{fig:current_systems}. 
We consider the results of each Group in turn.

\subsubsection{Group 1 systems}
Group 1 (left panel in figure~\ref{fig:frhab}) contains systems where the planet has a considerably larger semi-major axis than the outer edge of the HZ and a low eccentricity, or has a moderate eccentricity with 
a much larger semi-major axis. All of the systems show high values of $f_{\rm hab,1P}$: usually 
essentially unity. In 
all cases the mean value of $f_{\rm rhab}$ is lower and the distribution shows a broader range of values.

For the three systems with higher eccentricities (HD~72659, GJ~328 and HD~25015) the width of the distribution can be understood by considering figures~~\ref{fig:fracs} and~\ref{fig:fullpage}. The higher the observed
eccentricity, the larger the fraction of 3E runs contributing to the  $f_{\rm rhab}$-distribution. We have seen 
that 3E runs tend to be a lot more disruptive to test particle orbits. However, GJ~328 and HD~25015 differ from all the other Group 1 systems in that the giant is considerably more distant from the HZ, resulting in nearly all of the 4H runs along with a significant fraction of the 3E runs giving $f_{\rm rhab}\approx 0.9$

\begin{figure}
\centering
\includegraphics[width = 1\columnwidth]{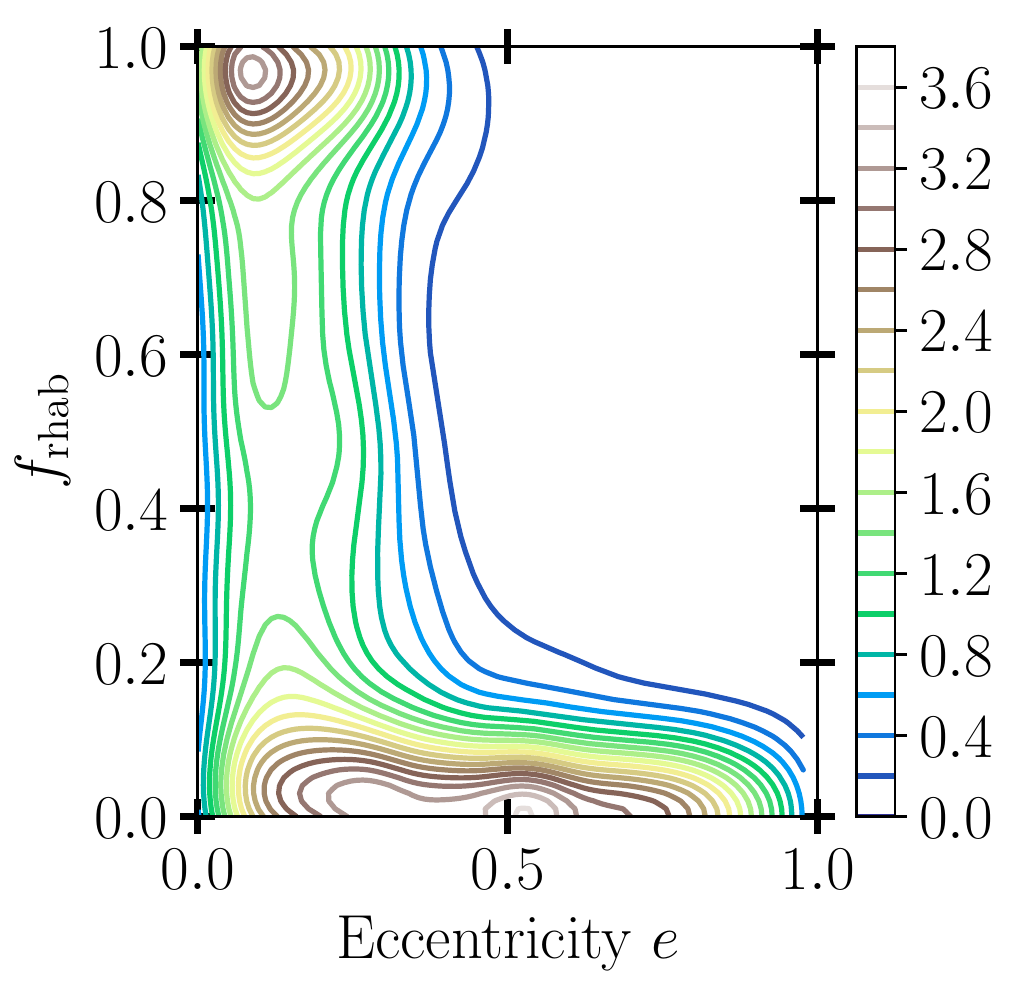}
\caption{A kernel density estimate showing the resulting $f_{\rm rhab}$-values for all our runs, both 3E and 4H, plotted against the final eccentricity of the planet.}
\label{fig:KDE}
\end{figure}

The other systems in Group 1 are all at low eccentricities and the $f_{\rm rhab}$-distribution is therefore dominated by 4H runs. The width of the $1\sigma$-range comes from the diversity of outcomes of the 4H runs, which can be seen in figure~\ref{fig:fullpage}. More specifically, for the 4H runs we find that 40\% of the runs have $f_{\rm rhab}\approx1$ with the rest of them fairly evenly spread between 0 and 1. Figure~\ref{fig:KDE} shows the distribution of $e_{\rm final}$ and $f_{\rm rhab}$ for all the runs. We see that low eccentricities have a mix
of low and high values of $f_{\rm rhab}$ whilst eccentricities above 0.5 give only very low values
of $f_{\rm rhab}$.  Some $17\%$ of runs yield $f_{\rm rhab}>0.9$ whilst $46\%$ of runs 
result in $f_{\rm rhab}<0.1$.
The distribution for an individual system can often deviate from the average of all
runs shown here. The eccentricity at which the peaks appear is set by the underlying eccentricity distribution resulting from the planet--planet scattering. A larger planet/star mass ratio shifts the peaks to higher eccentricities. The relative strength of each peak is set by the distance between the HZ and planet in terms of Hill radii. 4H runs in systems with distant planets such as HD 95872 nearly always have $f_{\rm rhab}>0.95$ and for the 3E runs such systems (and only those) have a fraction of runs showing high survival. Systems where the planet sits close to the HZ like HD 72659 do the opposite. The 4H runs are distributed fairly evenly between $f_{\rm rhab}=0$ and $f_{\rm rhab}=0.95$ and the 3E runs are always fully destructive.

\subsubsection{Group 2 systems}
Group 2 (middle panel in figure~\ref{fig:frhab}) contains systems where the planet has a 
semi-major axis just outside the outer edge of the HZ with a low to moderate eccentricity, 
or has a larger semi-major axis with a sufficiently high eccentricity to bring the pericentre very close to the HZ. Such planets are more destructive to objects
within the HZ. Other than HD 68402 all of the systems have a low $f_{\rm hab,1P}$ and even lower $f_{\rm rhab}$. The reason for this can be understood by considering figure~\ref{fig:KDE}. The high eccentricity systems only include runs from a region in the plot where there are no runs with high $f_{\rm rhab}$, whereas the moderately eccentric ones that sit closer to the HZ have 4H runs that are shifted to significantly lower $f_{\rm rhab}$ values as discussed in the previous subsection for the Group 1 systems.

The outlier, HD 68402, is a low-mass giant sitting very close to the HZ on a nearly circular orbit. In its present day orbit it does very little damage to the HZ; however, during the scattering any eccentricity fluctuations or changes to its semi-major axis easily destabilise test-particle orbits in the HZ. HD 68402 clearly demonstrates how much the history of a system can matter.

\subsubsection{Group 3 systems}

\begin{figure}
\centering
\includegraphics[width = 1\columnwidth]{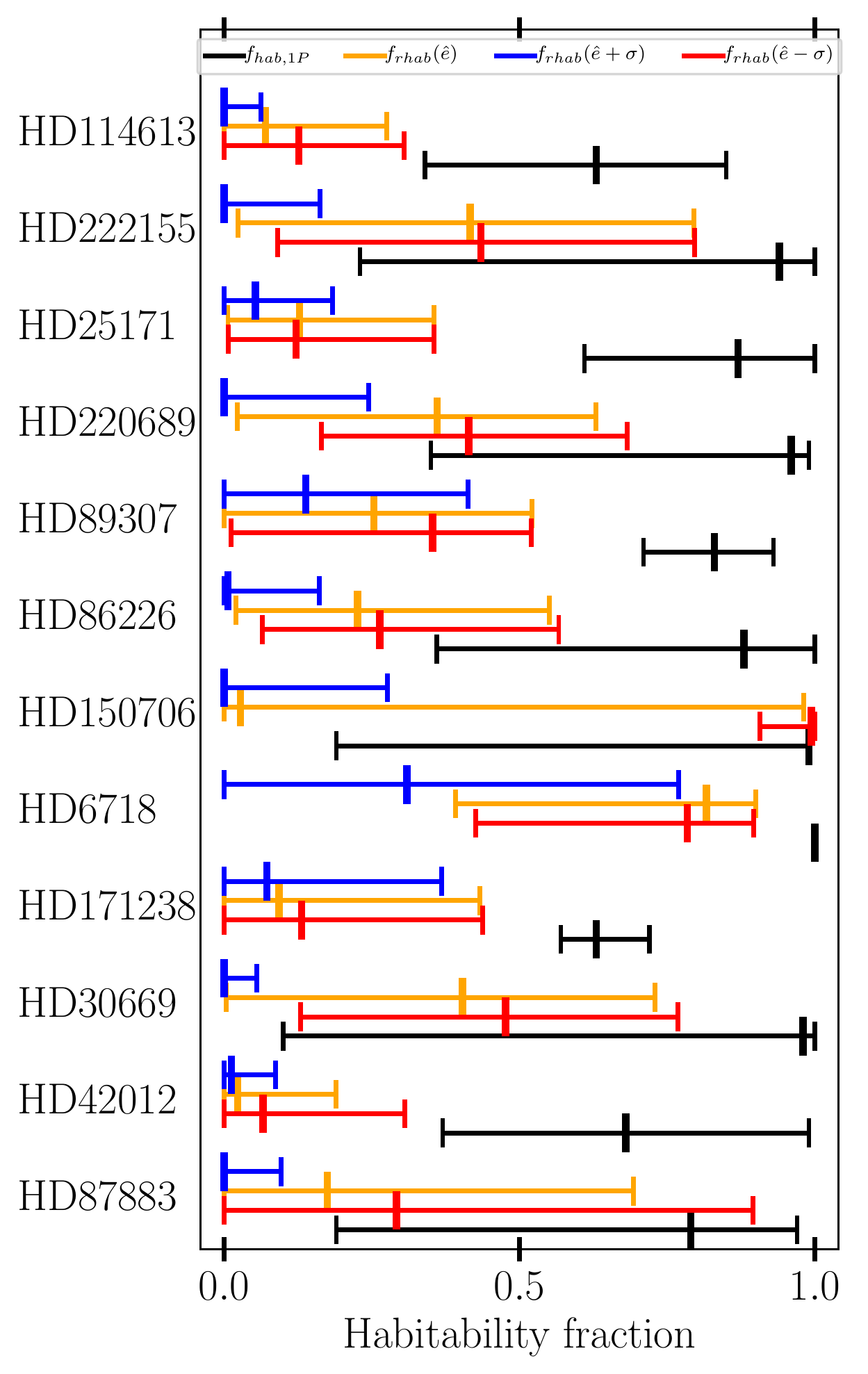}
\caption{The resilient habitability for systems with large uncertainties in 
  the eccentricity of the gas giant. The three $f_{\rm rhab}$ distributions 
  are generated using runs close to the $-\sigma$ (red), mean (orange) and 
  $+\sigma$ (blue) values of the eccentricity. 
  The black line shows $f_{\rm hab,1P}$, as in Figure~\ref{fig:frhab}.}
\label{fig:3lines}
\end{figure}

Group 3 contains the systems with a large uncertainty in the reported eccentricity of the planet and could therefore fall in either Group 1 or Group 2. Given that we include runs that are within one standard deviation of the reported eccentricity some time after the final planet ejection occurs, we are including runs from a much wider eccentricity range. We perform a second determination of the $f_{\rm rhab}$-distribution, in which we pick runs
close to the reported value of the observed eccentricity, and within $0.5\sigma$ of the $-1\sigma$ and $+1\sigma$ values (orange,
red and blue colours in figure~\ref{fig:3lines}).

The large eccentricity range means that we get a large sample of both 3E and 4H runs which widens the distribution. Not all distributions in figure~\ref{fig:3lines} are narrower than the corresponding ones in figure~\ref{fig:frhab}; however in a subset of them the majority of 3E runs get relegated to the high $e$ distribution (blue line in figure~\ref{fig:3lines}). This often shifts the other two distributions (orange and red lines) to higher values, in particular for HD 150706. If its actual eccentricity is close to the lower end it is one of the most resilient systems we have looked at. If it is at the high end, the opposite is true.

\section{Discussion}\label{sec:disc}

\subsection{Mechanisms responsible for the removal of particles from the HZ}\label{sec:ej}

\label{sec:resonaces}

We have shown that particles orbiting in a star's habitable zone 
can themselves be destabilised during an instability among 
its giant planets. We now look in more detail at the 
physical mechanisms responsible for removing particles from the HZ. 
This differs depending on whether the outer system 
is an equal-mass 3E configuration, or a hierarchical 4H configuration.

The 3E runs frequently give $f_{\rm rhab}\sim0$ as a $\sim$ Jupiter-mass planet 
is often scattered into the HZ. This results in most test particles being directly 
gravitationally scattered and then ejected from the system: around 80\% of the 
test particles removed from the HZ in the 3E runs have a close encounter in the HZ with 
one or more planets. 

In contrast, in the 4H runs jupiter only ever enters the HZ in 
systems where it is initially located very close to it. However, 
we still see $f_{\rm rhab}\approx0$ runs in systems where this is 
not the case. This happens in one of three ways:
\begin{itemize}
\item Saturn gets scattered into a HZ crossing orbit and ejects most of the test particles (rare).
\item Uranus or neptune gets scattered into a HZ crossing orbit and stays there for some time. This increases the eccentricity of the test particles which are later ejected by jupiter (less rare).
\item Resonances from jupiter inside the HZ excite the eccentricities of test particles which are later ejected by jupiter (common).
\end{itemize}

Examining the close encounter history of the test particles, we find that 
60\% of all the ejected test particles in the 4H runs never experience 
a close encounter with a planet that takes places within the HZ.
Rather, their eccentricities are excited by orbital resonances with the 
giant planets, after which they are typically removed by gravitational scattering 
once their apocentre brings them close to jupiter. 
\cite{Matsumura2013} and \cite{Carrera2016} identified 
secular resonances as being of importance in pumping 
up particle eccentricities during scattering amongst 
outer giant planets. Here, we show that for our systems 
mean-motion resonances (MMRs) with jupiter can be 
more significant. We discuss the effects of secular resonances 
in Section~\ref{sec:sechz}.

MMRs occur when the orbital periods of two bodies lie close to 
a ratio of integers. Here we consider $1:m$ and $2:m$ resonances 
between a test particle in the HZ and the jupiter. These occur 
where the semimajor axis of the particle in the HZ lies at 
$a=\left(1/m\right)^{2/3}a_\mathrm{J}$ or 
$a=\left(2/m\right)^{2/3}a_\mathrm{J}$, where 
$a_\mathrm{J}$ is the semimajor axis of the jupiter. 
Therefore, as the jupiter's semimajor axis moves 
erratically inwards as it scatters and ejects the 
outer planets, the mean motion resonances 
jump through the habitable zone.

We illustrate the effects of the passage of MMRs 
by comparing two runs from our 4H simulations 
of HD~72659. One run has a high and one a low 
$f_\mathrm{rhab}$. In neither run did a planet 
enter the HZ, and both end with the jupiter 
possessing a similar eccentricity.

\begin{figure}
  \begin{center}
  
  \begin{subfigure}[t]{\columnwidth}
    \centering
    \includegraphics[width=\linewidth]{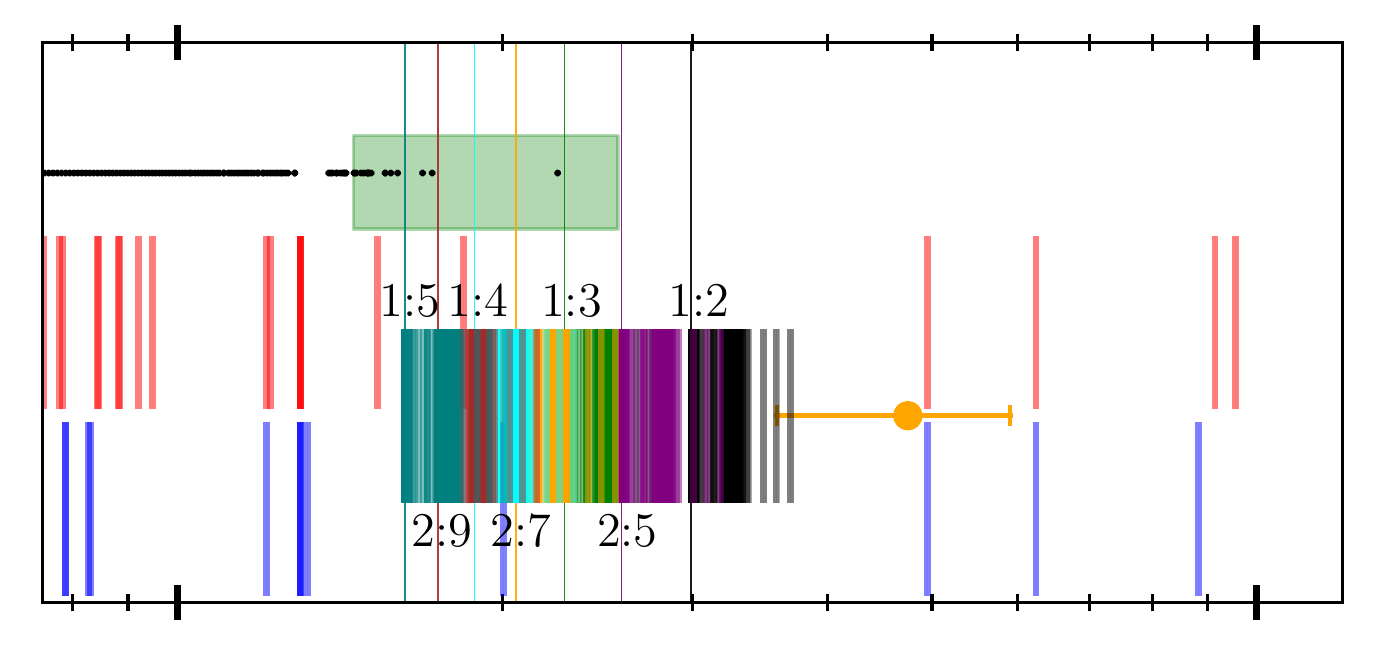} 
  \end{subfigure}
  \vspace{-.5 cm}
  \begin{subfigure}[t]{\columnwidth}
    \centering
    \includegraphics[width=\linewidth]{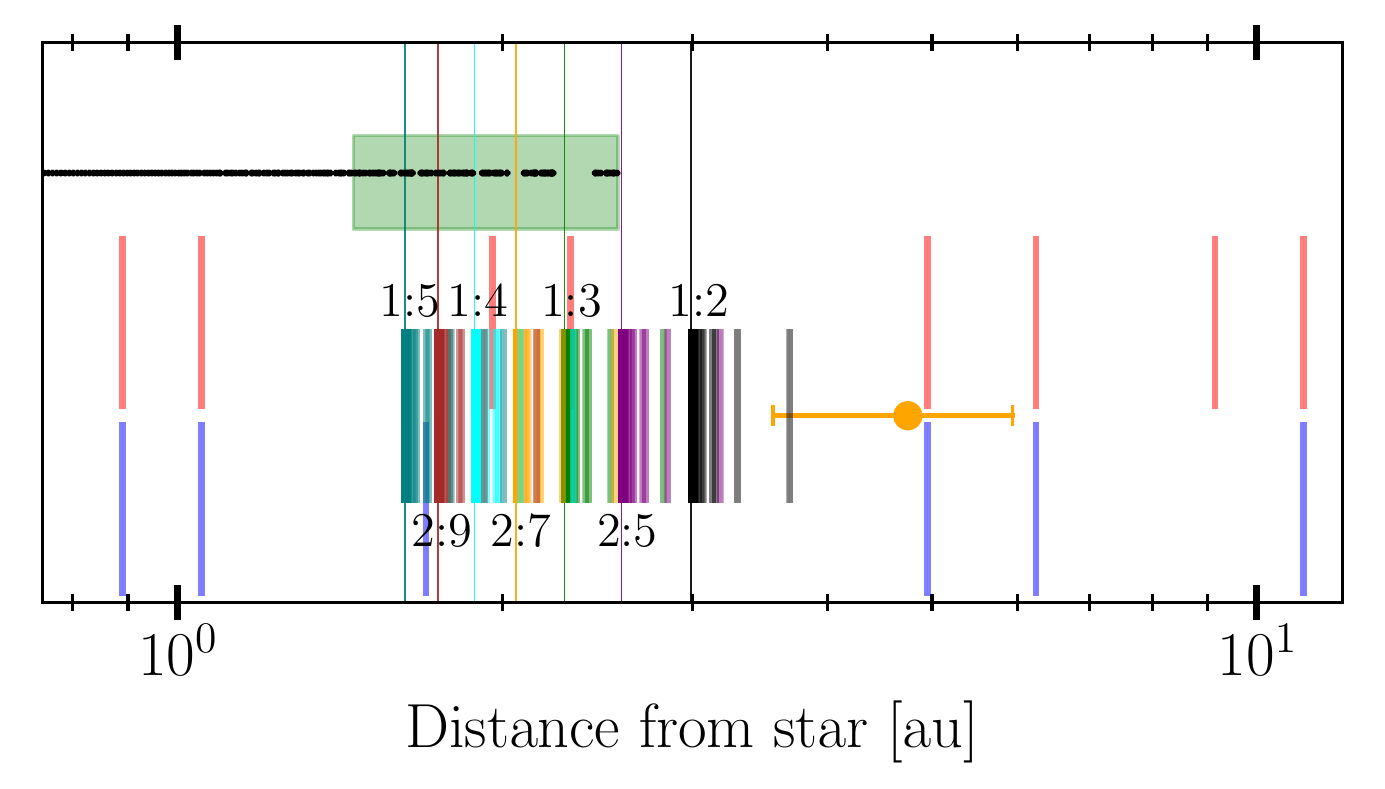} 
  \end{subfigure}
  \end{center}
	
  \caption{The figures show how the location of orbital resonances changes 
    with time for two different 4H runs of HD~72659. The final location 
    of the planet is shown in amber, the circle marking its semimajor 
    axis and the error bars its pericentre and apocentre; the habitable zone is shown in green, 
    and surviving test particles as small black points. Vertical lines mark the 
    location of orbital resonances, sampled every 10\,kyr.
    The off-axis red and blue lines show the location of the secular eccentricity 
    and inclination resonances. The remaining coloured lines show the $1:m$ and $2:m$
    mean motion resonances with the long line being placed at the final location of 
    the resonance.}
\label{fig:reses}
\end{figure}

The motion of the resonances in these two runs is shown in 
Figure~\ref{fig:reses}. Here the final location of the planet 
is shown in amber, and the HZ in green. Vertical lines 
mark the locations of MMRs and secular resonances, 
sampled every 10\,kyr, with the final locations of the MMRs 
marked as long vertical lines. The final location of each 
resonance lies at the innermost edge of its range over the 
integration, as expected from the net inward motion of the 
jupiter.

In the upper panel of Figure~\ref{fig:reses}, we illustrate a 
case where the jupiter moves inwards through a slow, 
incremental process. The sampled locations of the resonances 
lie dense in the habitable zone, meaning that particles 
at most semimajor axes in the HZ have time for their eccentricities 
to be highly excited by one or more resonances. Indeed, particles 
are found surviving the simulation at only a few semimajor axes, 
mostly towards the inner edge of the HZ. In the bottom panel, in contrast, 
the jupiter moves inwards more rapidly, meaning that a significant 
fraction of the HZ does not become excited by the passage of MMRs, and 
many particles survive the integration. 
These two examples suggest a dependence of a system's 
resilient habitability $f_\mathrm{rhab}$ on the nature of the 
motion of the jupiter.

\begin{figure}
\centering
\includegraphics[width = 1\columnwidth]{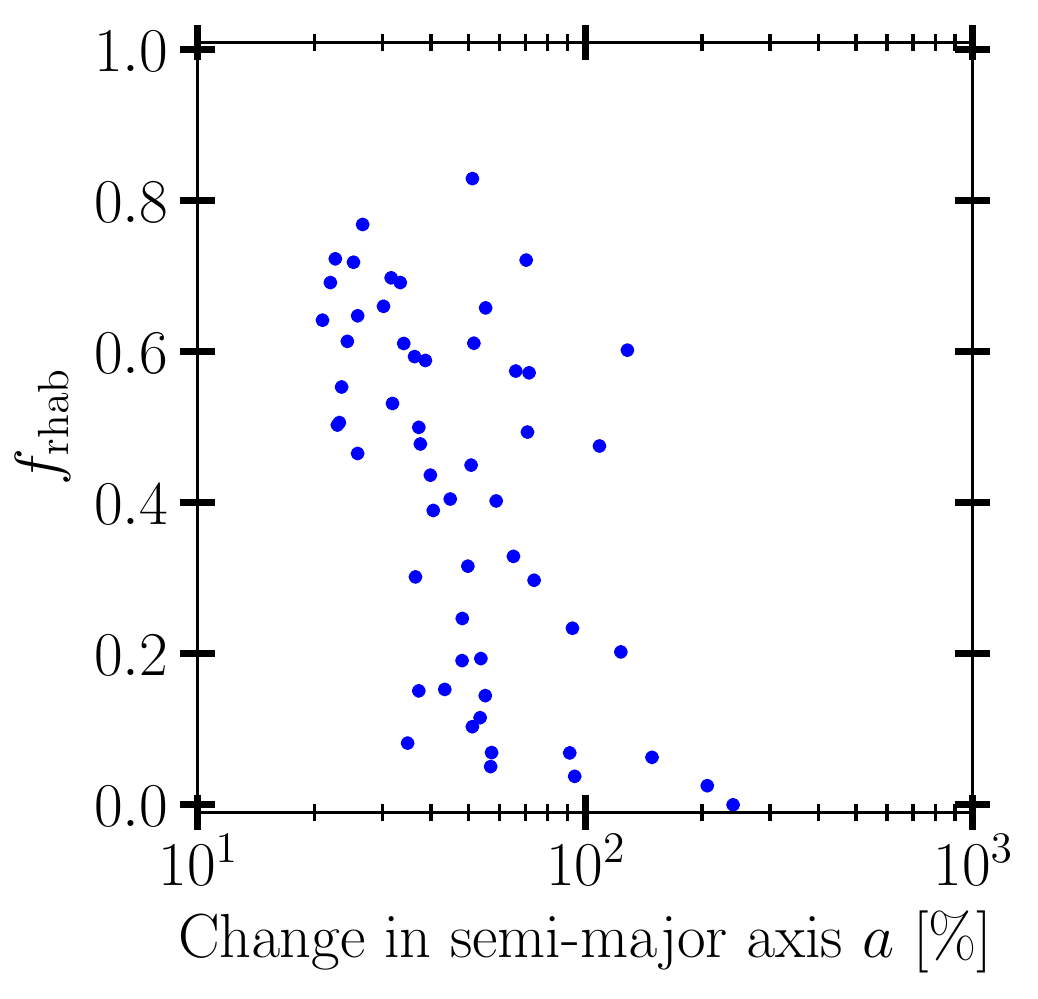}
\caption{$f_{\rm rhab}$ as a function of the summed up absolute fractional change 
  in semi-major axis for 
  all 65 of the the 4H-runs of HD~72659 where no planet enters the HZ. The 
  change in semimajor axis is defined by eq~\ref{eq:da}.}
\label{fig:f_v_a}
\end{figure}

We have explored a number of different parametrizations to correlate the evolution of jupiter's semi-major axis with $f_\mathrm{rhab}$. We find that looking at the fractional absolute change in semimajor axis between each output time (10\,kyr resolution) shows this dependence best. The fractional absolute change is given in the form
\begin{equation}
\frac{\Delta a}{a} = \sum_i \frac{|a_{i+1}-a_i|}{a_i},
\label{eq:da}
\end{equation}
\noindent where $a_i$ is the semi-major axis at the $i$th output. Thus, 
a jupiter that attains its final semimajor axis in a few large 
jumps will have a small $\Delta a/a$, while those that move in 
many small steps, which often involves temporary reversals of 
direction, will have a large $\Delta a/a$. We plot in Figure~\ref{fig:f_v_a}
the resilient habitability for HD~72659 against $\Delta a/a$, for all 
4H runs where no planet entered the HZ. As expected, we see an anticorrelation 
between $f_\mathrm{rhab}$ and $\Delta a/a$.

Combining the anticorrelation we see in figure~\ref{fig:f_v_a} 
with the fact that 60\% of ejected test particles in 4H runs never 
have a close encounter inside the HZ, we conclude that the 
majority of test particles in the 4H case are ejected due to 
MMRs sweeping over the HZ. 
The ejection efficiency is correlated with how much the semi-major 
axis of jupiter changes.

\subsection{Secular resonances in the habitable zone during and after an instability}\label{sec:sechz}

One way through which eccentricities can be forced to high 
values in the absence of direct scattering is through secular resonances. 
Secular resonances occur when the precession frequency of one of the 
particles in the HZ matches one of the secular eigenfrequencies 
of the giant planets. These resonances occur at specific values of 
the particles' semimajor axis, and formally the forced eccentricity 
is infinite at these locations. Because the planetary secular frequencies 
are dependent on their semimajor axes, the frequencies change during 
planet--planet scattering, and so the secular resonances 
also move. This process was identified by \cite{Matsumura2013} and 
\cite{Carrera2016} as a means of destabilising terrestrial planets during 
an instability amongst giant planets.

Returning to Figure~\ref{fig:reses}, we have plotted the location 
of the secular resonances (both eccentricity and inclination) 
in the two runs every 10\,kyr. The resonance locations are calculated 
according to linear Laplace--Lagrange theory \citep[][Ch.\ 7]{Murray1999}. 
They move inwards as the planets' semimajor axes change during scattering, 
similar to the motion of the MMRs, but in contrast to the MMRs they move 
much further and more rapidly. This is because they depend on the semimajor 
axes of all planets, not just the jupiter; but the jupiter alone sets the location 
of the MMRs and itself moves comparatively little. Indeed, the secular resonances 
pass across the whole HZ in just a few 10s of kyr, insufficient time to 
significantly affect particles' eccentricities. Typically the secular resonances 
pass through the HZ in less than 30\,kyr, and only in $\sim5\%$ of runs do 
they spend more than 100\,kyr in the HZ. In the upper panel we do, 
however, see that the secular eccentricity and inclination resonances remain 
for several 10s of kyr just interior to the HZ, where they destabilise a 
few bodies and knock a small hole in the distribution of survivors. 

\begin{figure}
\centering
\includegraphics[width = 1\columnwidth]{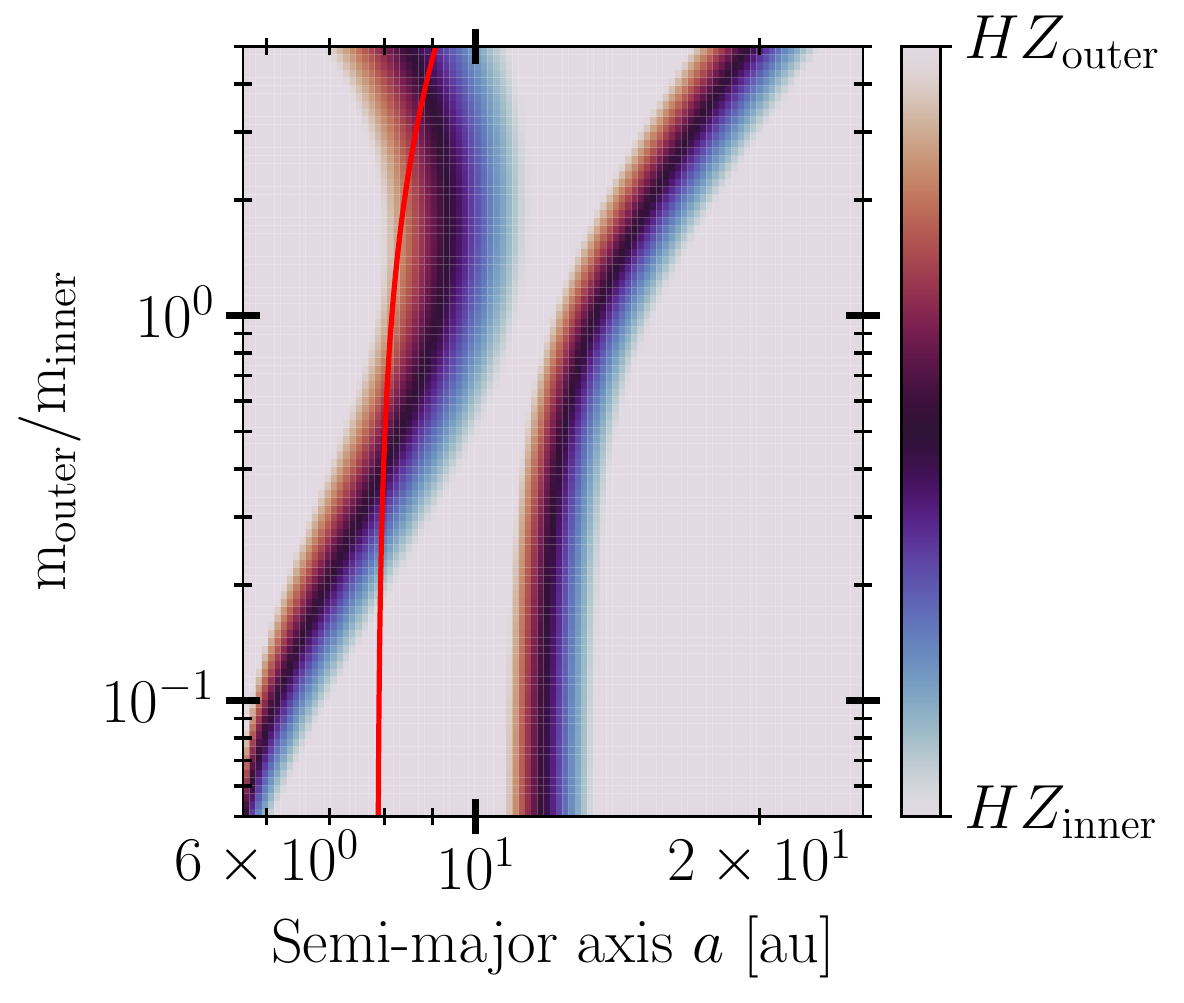}
\caption{For the system HD~95872 ($a=4.75$\,au) the figure shows at what 
  semi-major axis a planet of a given mass needs to be placed for either of the secular 
  eccentricity resonances to be in the HZ. The colour indicates where in the 
  HZ the resonance appears. The red line shows the Gladman stability limit for circular orbits: 
  systems wider then this cannot experience orbit-crossing.}
\label{fig:sechz}
\end{figure}

While in principle the final giant planet configuration following scattering 
could leave a secular resonance in the HZ, this is very rare, occurring 
in $\lesssim1\%$ of our runs. The reason is that the planets must be spaced 
rather close to each other to place a secular resonance in the HZ. We 
show in Figure~\ref{fig:sechz} where the secular eccentricity resonances would lie in 
the HD~95872 system if it contained a second giant planet, as a function 
of this planet's mass and semimajor axis. To place a secular 
resonance in the HZ requires a separation of $\sim4-7$ mutual Hill radii 
between the planets. These configurations would be stable if the planets' 
orbits were near-circular, as they satisfy the limit for Hill stability 
given by \cite{Gladman1993}:
\begin{equation}
\frac{a_{\rm outer} - a_{\rm inner}}{R_{\rm MHill}}>2\sqrt{3},
\end{equation}
where $R_{\rm MHill}$ is the mutual Hill radius 
from equation~\ref{eq:hill} and $a_{\rm outer},a_{\rm inner}$ are 
the semi-major axes of the planets. We mark this limit in Figure~\ref{fig:sechz}
as a solid red line. While the circular configurations would be stable, scattering often 
leaves planets with significant eccentricities, and this significantly 
increases the separation required for the planets to remain 
stable after scattering 
\citep[e.g.,][]{MardlingAarseth01,MustillWyatt12,Giuppone2013,Petrovich15,Antoniadou2016,Hadden2018}. 

\subsection{Low-eccentricity systems}\label{sec:low_e}

For the low-eccentricity systems we have to consider whether 
or not the system has been through a phase of planet--planet 
scattering, and in either case, what the value of $f_{\rm rhab}$ will be.
If the system has undergone an instability, then the answer is given in figure~\ref{fig:frhab}. 
If there has not been any scattering the only way in which $f_{\rm rhab}$ could be significantly different from $f_{\rm hab,1P}$ is if there is an additional hitherto undetected planet in the system. This can affect the system either through changing the eccentricity of the observed planet over time, or by giving rise to secular resonances inside the habitable zone as we have just discussed. 

\subsubsection{Eccentricity changing with time}

Secular interactions lead to eccentricities varying with time. These variations occur on timescales of thousands of years, therefore the $f_{\rm hab,1P}$-simulations do not necessarily capture the full picture. In particular, a planet with a low eccentricity at 
the present day may be at the low point of a secular cycle, if there 
is an undetected exterior planet currently on an eccentric orbit.

We test how damaging this can be by determining what we call 
$f_\mathrm{hab,2P}$: here, we take systems where two planets survive 
our 4H runs and the jupiter has an eccentricity consistent 
with the observed system, repopulate the HZ with low-eccentricity test particles, 
and run the system for 10\,Myr. $f_\mathrm{hab,2P}$ is then the 
fraction of these repopulated test particles that survive the 
integration. By choosing two-planet systems from among 
our post-scattering configurations, where the outer planet 
often has a highly-eccentric orbit, we are generating systems 
where we might expect that a large eccentricity can be passed to 
the inner planet through secular interactions, and so where 
$f_\mathrm{hab,2P}$ will be low.

However, what we find is that in most cases the variation of the 
inner planet's eccentricity is smaller than its observational 
uncertainty, and that the $f_{\rm hab,2P}$ values lie 
mostly within the range of $f_{\rm hab,1P}$ in Figure~\ref{fig:frhab}.
$f_\mathrm{hab,2P}$ is lower if the mean eccentricity of the jupiter 
over its secular cycles is higher than the reported value, and lower if the 
reverse is true. In a few percent of cases, the eccentricity of the jupiter at the 
end of the original integration (the start of the repopulated integration)
is close to the minimum or the maximum in its cycle, and in these cases, 
$f_{\rm hab,2P}$ can differ significantly from $f_{\rm hab,1P}$. 

We now return to the focus of this subsection: low-eccentricity 
planets that have not undergone instability. 
The amplitude of the inner planet's eccentricity oscillations is 
larger if the outer planet's eccentricity is larger, and so 
systems that have not undergone instability will 
usually have values of $f_\mathrm{hab,2P}$ similar to 
their values of $f_\mathrm{hab,1P}$, since this is usually the 
case for more dynamically excited systems too.

\subsubsection{Placing the secular resonances in the habitable zone}

Nevertheless, there is a potential exception to the 
fact that systems that have never undergone planet--planet 
scattering will have a high $f_\mathrm{hab,2P}$: this is 
if they have secular resonances in the habitable zone. We test the effect such 
secular resonances can have by setting up runs for the HD~95872 and 
HD~72659 systems, adding a planet of equal mass on a circular orbit
at a semimajor axis that places a secular resonance in the HZ,
populating the HZ with test particles, and integrating the system 
for 10\,Myr, verifying that the two planets are stable.

For HD~95872 we find that $f_{\rm hab,2P}$ is as high as 
$f_{\rm hab,1P}$ when we place a secular resonance in the HZ; 
however, a large fraction of the test 
particles in the HZ acquire high eccentricities above $0.5$. 
As we discuss in Section~\ref{sec:ecc_effect}, these high eccentricities
are likely detrimental to any planet's habitability.
The secular resonance in this system is especially strong 
because of the large planet mass relative to the 
star ($3.74\mathrm{\,M_J}$ versus $0.7\mathrm{\,M_\odot}$). 
In HD~72659, the test particles that have large eccentricities 
excited eventually get ejected, since the giant planet's 
pericentre is close to the HZ. HD~95872 and HD~72659 show 
respectively the highest and the lowest values of both 
$f_\mathrm{hab,1P}$ and $f_\mathrm{rhab}$, showing that 
the effects of secular resonances could be significant 
for many systems.

As discussed above, it is less likely that systems with high-eccentricity planets 
have secular resonances in the HZ because of the constraints imposed 
by stability of the gas giants. In answering the question ``how 
likely is a habitable zone to be destabilised by a secular resonance?'', 
we therefore have a curious inverse dependence on eccentricity, 
where the systems with more eccentric giant planets are less likely 
to have their HZ destabilised by a secular resonance. This does, 
however, ignore the scattering history of the giant planets, 
which may have destabilised bodies in the HZ by scattering and the passage of MMRs.

\subsection{The effect of eccentricity on habitability}

\label{sec:ecc_effect}

The habitable zone is defined as the spherical shell around a star at
which the flux received allows for liquid water to exist on the surface of the
planet under a set of atmospheric conditions. In our simulations we have
hitherto considered only whether the semi-major axis remains in the HZ, ignoring
any orbital eccentricity the planet may possess.

At some point this approximation breaks down and we have to consider the
effects of having large temperature differences between peri- and apo-centre.
In general, the higher the orbital eccentricity at a given semimajor axis,
the higher the orbit-averaged radiation flux from the star: therefore,
the habitable zone is more distant at higher eccentricity.
While simple energy-balance arguments give a dependence
of the HZ on eccentricity of $a_\mathrm{HZ}\propto\left(1-e^2\right)^{-1/4}$,
more detailed modelling \citep[e.g.,][]{Dressing2010,Spiegel2010,
Linsenmeier2015,Bolmont2016,Kane2017}, incorporating the effects
of planetary spin rate and obliquity, shows that high eccentricities ($e\gtrsim0.5$)
can have a significantly stronger effect.

\begin{figure}
\centering
\includegraphics[width = 1\columnwidth]{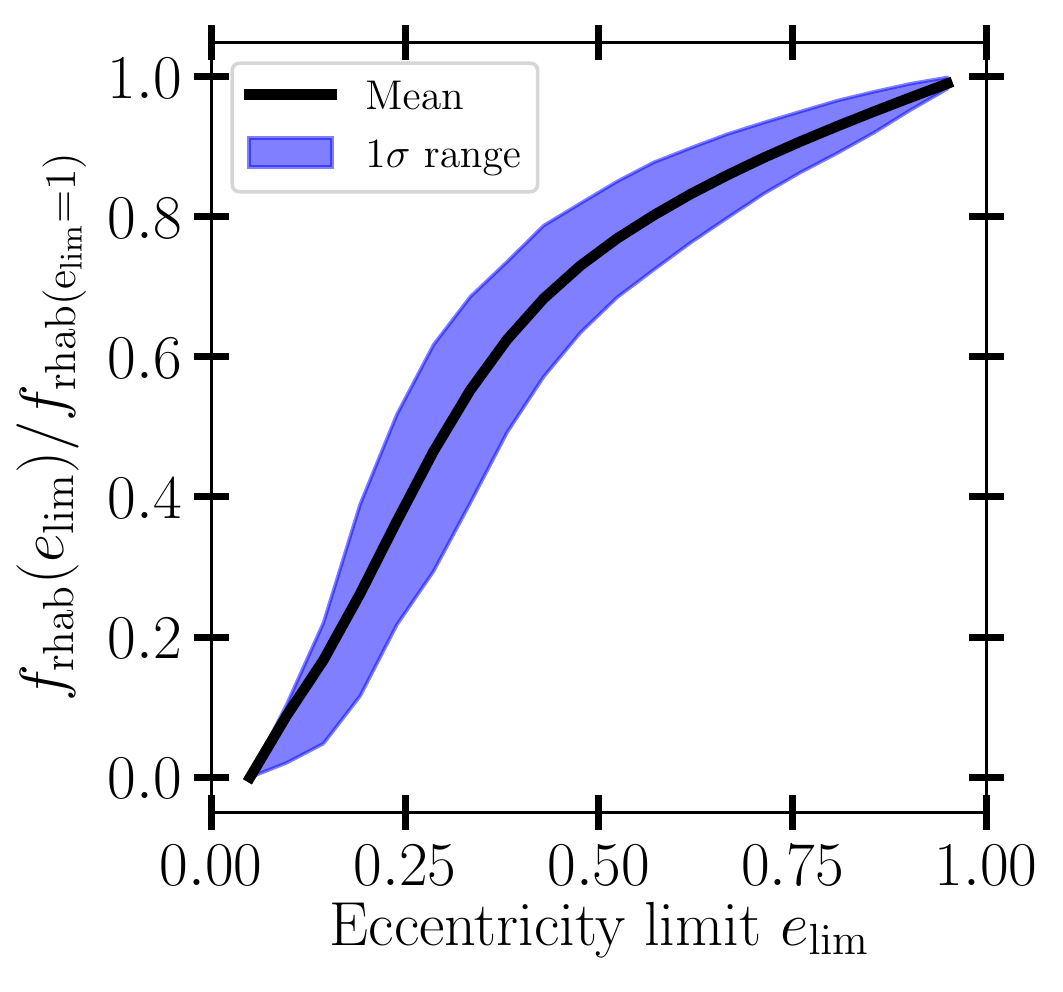}
\caption{The change in $f_{\rm rhab}$ for all systems when an eccentricity limit is
  imposed on the habitability. For each value of $e_{\rm lim}$ we recalculate
  $f_{\rm rhab}$ and do not count any test particle toward it that
  reaches an eccentricity greater than $e_{\rm lim}$ throughout the run.
  The $1\sigma$ error band shows the standard deviation over all runs
  of all systems.}
\label{fig:ecrit}
\end{figure}

As the actual effect of eccentricity
depends on other parameters such as the spin rate and obliquity, we
explore the number of stable particles that attain a specified value
of eccentricity, $e_\mathrm{lim}$, at any point in the run.
We then recalculate the resilient habitability $f_\mathrm{rhab}$
by excluding the particles that reached a high eccentricity, assuming
that they have been rendered uninhabitable. By performing this
calculation for a range of $e_\mathrm{lim}$, we can find how
significantly the inclusion of an eccentricity consideration
affects $f_\mathrm{rhab}$, without committing ourselves to a single
(model-dependent) eccentricity limit. We then show in Figure~\ref{fig:ecrit}
the ratio of $f_\mathrm{rhab}$ with the eccentricity limit to that
without. As expected, this rises from zero at $e_\mathrm{lim}=0$ (all
particles are considered uninhabitable when they acquire any eccentricity)
to unity at $e_\mathrm{lim}=1$ (all surviving particles are considered
habitable, regardless of eccentricity). For $e_\mathrm{lim}=0.5$,
about 30\% of surviving particles are rendered uninhabitable by
acquiring eccentricities above $e_\mathrm{lim}$. This is a significant
reduction in resilient habitability, motivating further
studies of the climate of highly-eccentric terrestrial planets.

\subsection{Additional considerations}

\subsubsection{Multiple massive terrestrial planets}

We have used massless test particles to represent terrestrial 
planets in our simulations. This is computationally efficient 
as one can simulate many hundreds of test particles within one 
single run. The total mass of the terrestrial planets 
is $\lesssim1\%$ of that of the giant planets, so using test particles
to represent massive terrestrial planets should not be a problem 
when considering the direct interactions between the particles 
and the giant planets. However, use of massless non-interacting test 
particles means that we do not account for any interactions 
between the habitable planets, should more than one be present.

In principle, if there are multiple terrestrial planets close to each other 
they can couple to each other and counteract external perturbations. The 
stabilization by such couplings has been confirmed for both the Solar System (see e.g. ~\citealp{Innanen1997, Batigyn2008, Laskar2009, Zeebe2015}), 
and generally for tightly packed systems (see 
e.g.~\citealp{Malmberg2011, Kaib2011, Hansen2017, Mustill2017, Denham2019}). 
The key factor determining whether the system is stabilised is the time-scale. 
If the coupling between the planets is on a short time-scale compared to 
the perturbation the planets will be resistant to it. Thus, we might expect 
the resilient habitability of systems to increase if the test particles are 
replaced with a handful of massive planets.

On the other hand,~\cite{Carrera2016} investigated the resilient habitability 
of systems where the habitable zone was populated either by a swarm of 
test particles, or by one, two or four massive ($1\mathrm{\,M}_\oplus$) planets. 
As expected, the survival rate of the single planets is consistent with that 
of the test particles, but the survival of HZ planets in the systems 
of two or four terrestrial planets is in fact lower than that of the test particles. 
In practise then, multi-planet systems subject to strong scattering from 
an outer system are slightly more vulnerable and more likely to be destabilised, 
although in systems with weaker external forcing it is likely that 
the protective coupling can still stabilise the system.

\subsubsection{The timing of the instability}

It is thought that terrestrial planets usually take $10-100$\,Myr until they are fully formed.
This time-scale is typical in simulations of terrestrial planet formation from swarms 
of planetary embryos \citep{Chambers2001,OBrien2006,Hansen2009,Raymond2009,Lambrechts2019}, 
and agrees with radiogenic dating of the Earth and Moon \citep{Jacobsen2005,Halliday2008}. 
On the other hand, gas giants must form in a few Myr while the gas disc is still present 
\citep{Machida2010,Piso2014,Piso2015,Bitsch2019}. The instability amongst the giant planets 
occurs some time after they have formed. It is a steep function of their orbital 
separation, and the time-scale of onset at a given separation can vary by an 
order of magnitude~~\citep{Chambers1996, Shikita2010}. Therefore, the instability can 
take place before or after terrestrial planet formation is complete. Indeed, in 
consistent simulations of formation, migration and dynamics, \cite{Bitsch2019} found 
that 5\% of the giant planet systems they formed were unstable on timescales $<10$\,Myr.

Thus, depending on the timing of the instability, the terrestrial planets may be fully formed or may still be a swarm of embryos. Whether the test particles in our simulations are considered to represent fully-formed planets or planetary embryos leads to two different interpretations.

If the instability happens after terrestrial planet formation has finished, then each particle lost from the simulations represents one of a handful of fully-formed planets. Each run can then be interpreted as simulating a large number of different realisations of an inner terrestrial planet subject to the same forcing from the unstable gas giants. Thus, a value of $f_{\rm rhab}$ of $25\%$ simply means that on average one quarter of planets would be retained and three quarters lost. If multiple fully-formed terrestrial planets are present in the same system, then their mutual interactions can affect the dynamics: \cite{Carrera2016} showed that this results in a lower fraction of planets surviving.

On the other hand, if the instability occurs before terrestrial planet formation is completed, our test particles can more readily be interpreted as representing the swarm of planetary embryos present in the habitable zone in one system. Thus, a value of $f_{\rm rhab}$ of $25\%$ now means that the habitable zone loses three quarters of its embryos. Depending on their distribution and excitation, terrestrial planet formation may not be able to proceed all the way to Earth mass, but may stall at smaller masses as much of the available material can be ejected from the system during the instability. An additional complication concerns the outcome of collisions between embryos: when excited by an instability in the outer system, collision velocities increase and collisions between embryos result in significant mass loss \citep{Mustill2018}. This can reduce the final size of the terrestrial planets still further, if the ejecta is ground down and removed by radiation pressure before being reaccreted by one of the embryos.

\subsubsection{RV mischaracterization of planet eccentricity}\label{sec:RV}

The signal from a planet with a seemingly eccentric orbit may, 
in some circumstances, actually be generated by a planet on a 
circular orbit. There are two ways in which this could happen.
The first arises from the fact that eccentricity is a positive-definite 
quantity: it cannot be negative. Therefore, the fitted eccentricities 
for planets on circular or low-eccentricity orbits are biased towards higher
values~\citep{LucySweeney71,ShenTurner08,Zakamska2011,Hara2019}. 
For example, \cite{Zakamska2011} show that the fraction of low eccentricity 
systems could be underestimated by as much as a factor of 3. This 
means that the true values of the resilient habitability for some of 
our systems may be higher than we have calculated, as the known planets' 
eccentricities may be somewhat smaller than the reported literature values.

The second way in which a seemingly eccentric signal can be generated 
by a planet on a circular orbit is through the presence of an interior, lower mass 
planet at a 2:1 period ratio. \cite{Anglada2010} showed that two such planets 
show a similar and sometimes indistinguishable RV signal as only having the 
outer, more massive planet on an eccentric orbit. They show that the RV 
signal from as many as 35\% of observed eccentric single planet systems show no 
statistically significant difference from being two planets in a 2:1 configuration 
rather than a single eccentric planet. 
Further, \cite{Wittenmyer2019} sets an upper bound on the possible apparent eccentricity at 0.5.

We first make a simple estimation of the destructive effects on the HZ of 
a single eccentric planet compared to two planets in a 2:1 configuration. 
We estimate the mass of the inner companion required to mimic 
an eccentric orbit from figure~3 of \cite{Kurster2015}, and compare the 
potential for direct scattering in both configurations, comparing 
the reach of the Hill radius of the inner planet $a_\mathrm{in}-r_\mathrm{H,in}$
with that of the seemingly eccentric outer planet at pericentre
$a_\mathrm{out}(1-e_\mathrm{out}) - r_\mathrm{H,out}$. We find that 
the eccentric outer planet will be more effective at direct scattering 
than the 2:1 two-planet configuration.

We test this with a set of simulations and find that systems with 
$e<0.3$ also can be more damaging than their 2:1 equivalent. This occurs 
if the pericenter distance to the outer edge of the HZ is smaller than the 
distance from the extra interior planet to the edge in terms of Hill's radii. 
This is the case for nearly all systems in this range that are fairly close to the HZ, 
barring the ones where the 2:1 resonance sits inside the HZ which already show low 
$f_{\rm hab,1P}$ and mostly $f_{\rm rhab}=0$. The most resilient and distant 
systems such as HD~95872 show no difference when replaced with the 2:1 equivalent.

\subsubsection{Planet masses and multiplicities}

We now consider two complications to our treatment of the planet masses 
in our simulations. First, we have set the planet masses in the simulations to 
equal $M\sin I$ of the observed planet, which on average will underestimate the 
planet mass by a factor of 1.6, and is a strict lower limit on the mass. To 
explore the effect of the true mass being higher, we have re-run the 
single-planet simulations (to calculate $f_{\rm hab,1P}$) for 
HD~95872 and HD~72659 with planetary masses increased by factors of 
1.4, 1.7 and 2.0. For HD~95872 there is no difference in the outcome. 
For HD~72659, where the giant's pericentre is closer to the HZ, we find a small 
reduction in $f_{\rm hab,1P}$ of $0.02-0.06$. 
A larger mass is marginally more disruptive if the pericentre is within a few 
Hill radii of the HZ, but because the Hill radius is a weak function of mass (Equation~\ref{eq:hill})
the effect is not very strong.

When considering $f_{\rm rhab}$, we expect that the peak of the eccentricity 
distribution after scattering will shift to slightly higher values with increasing planet--star mass 
ratio. However, we also expect that the mass ratio \emph{between} planets in the system is more 
significant in setting the final eccentricity distribution and the history 
of scattering. We note here that soon \emph{Gaia} will provide a measure of the true 
mass of many Jupiter analogues \citep{Sozzetti2008,Perryman2014,Ranalli+18}, 
permitting a refinement to the estimation 
of $f_{\rm hab,1P}$ and $f_{\rm rhab}$ by repeating these simulations with the real planetary masses.

The second complication is that we have reduced the set of 
planetary mass ratios considered to two choices: strictly equal masses (3E), 
or a hierarchical Solar System analogue (4H). Although we have shown that this simplification 
can reproduce the observed eccentricity distribution of the giant exoplanets, 
in reality mass ratios will fall on a continuum; this greatly increases the parameter 
space to investigate. Here, we simply note the following: 
With our dichotomous distribution of mass ratios, for 
a given eccentricity the distribution of $f_{\rm rhab}$ is bimodal (figure~\ref{fig:KDE}), 
having a high-survivability peak from the 4H systems and a low-survivability peak 
from the 3E systems. We would then expect, for example, a 2:1:1 mass 
ratio to be intermediate between the 4H and the 3E
in destructibility, and
with a continuum of mass ratios, we expect the $f_{\rm rhab}$ distribution to 
be less bimodal.

Finally, the number of planets in our simulations was 
fixed at either three or four. Adding more planets will likely be more destructive 
to bodies in the HZ if the planets are comparable in mass to the largest planet, 
since there will be more possibilities of scattering a large planet onto an orbit with a 
small pericentre, and also more secular resonances to destabilise the HZ at a distance. 
However, if the additional planets are smaller (neptune-mass), there should be a weaker effect, 
as we have shown that the dominant mechanism for destabilising 
the HZ is not the entry of neptune-mass planet into it (Section~\ref{sec:ej}).

\section{Future observations}\label{sec:obs}

\begin{figure}
\centering
\includegraphics[width = 1\columnwidth]{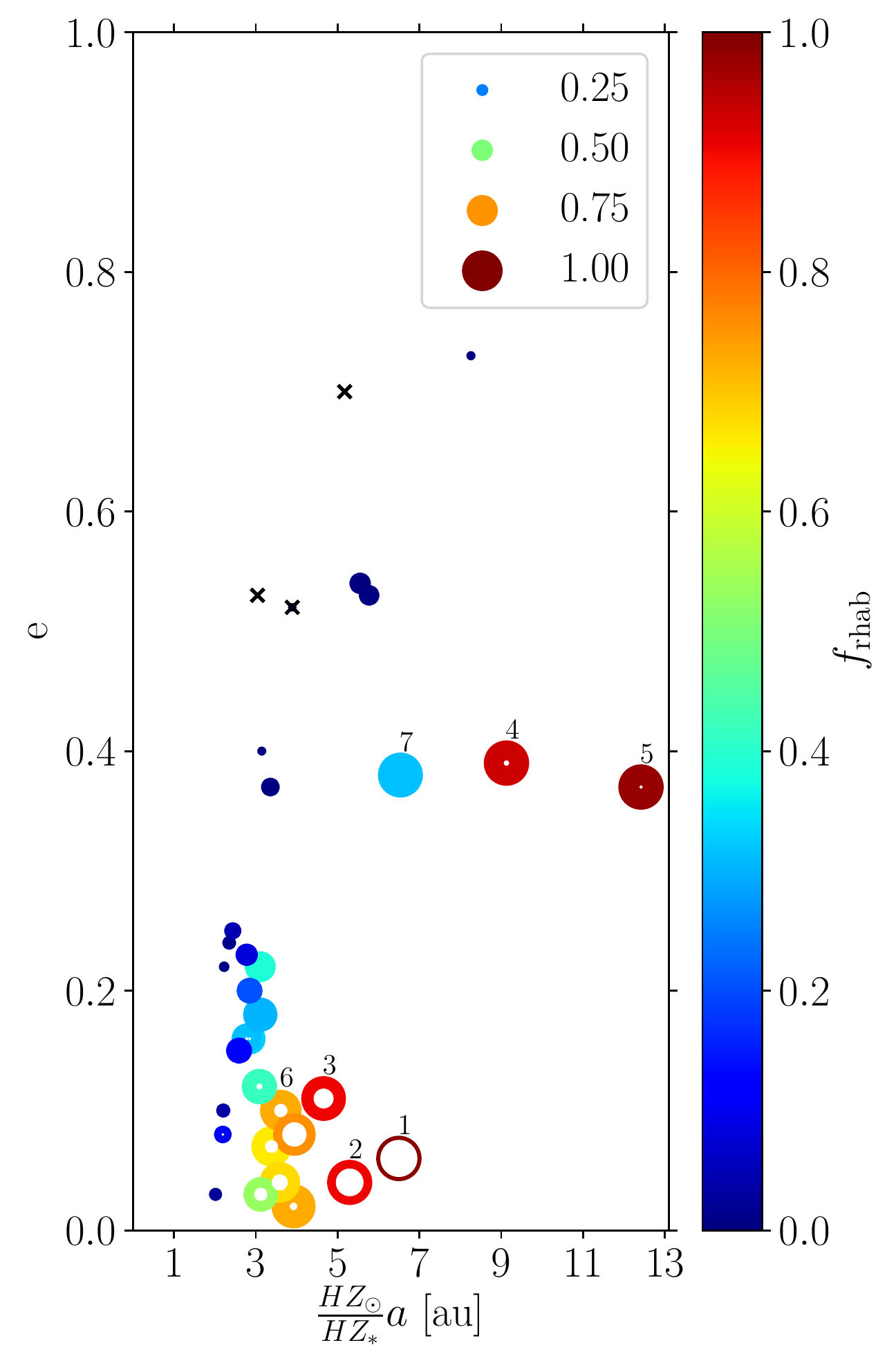}
\caption{The figure shows the resilient habitability, and its uncertainty, of all 
  the simulated systems. The systems are marked with their planets' 
  semimajor axes rescaled so that the midpoint of the systems' HZ corresponds 
  to the midpoint of the HZ in the Solar System, and with their reported 
  eccentricities. The colour of each symbol is set by the median of 
  $f_{\rm rhab}$-distribution. The inner and outer radii of annular symbols scale 
  linearly with the $\pm1\sigma$ values of $f_{\rm rhab}$-distribution. The crosses 
  indicate systems where the $1\sigma$ upper limit is  less than $0.05$. 
  Thus, the blue crosses are very bad, small blue dots are bad, 
  the narrow brown/red annuli are good and the wide annuli give a range of 
  possible outcomes.
  We have have highlighted the best potential systems for follow-up observations; 
  they are as follows:\\
1) HD~95872 2) HD~154345 3) HD~102843 4) HD~25015 5) GJ~328 6) HD~6718 7) HD~150706.}
\label{fig:obs}
\end{figure}

Finally, we can make some comments as to which of the systems we have 
studied provide the best prospects for the detection of planets in the 
HZ, based on their resilient habitability. All the low-eccentricity systems 
in Group 1 make good candidates for observations as they all have 
high median $f_{\rm rhab}$-values. Particularly good are the ones where the 
lower end of the orange interval in Figure~\ref{fig:frhab} is high: HD~95872, 
HD~154345 and HD~102843. These are highlighted in Figure~\ref{fig:obs}, 
where we show the resilient habitability for all of the systems we have 
studied as a function of the giant planet's eccentricity and 
distance to the HZ. This figure should be compared to figure~12 of \cite{Carrera2016}, 
which shows the median resilient habitability for a generic system. The 
broad features are the same -- high resilient habitability for low eccentricity 
and wide orbit planets -- but we wish to emphasize two points: the 
large range in resilient habitability arising from different runs 
in the same system; and the fact that different systems close to each other 
in parameter space on this Figure can have different values of 
resilient habitability, in part because we have included the mass 
of the known planet as a third parameter.

\begin{figure}
\centering
\includegraphics[width = 1\columnwidth]{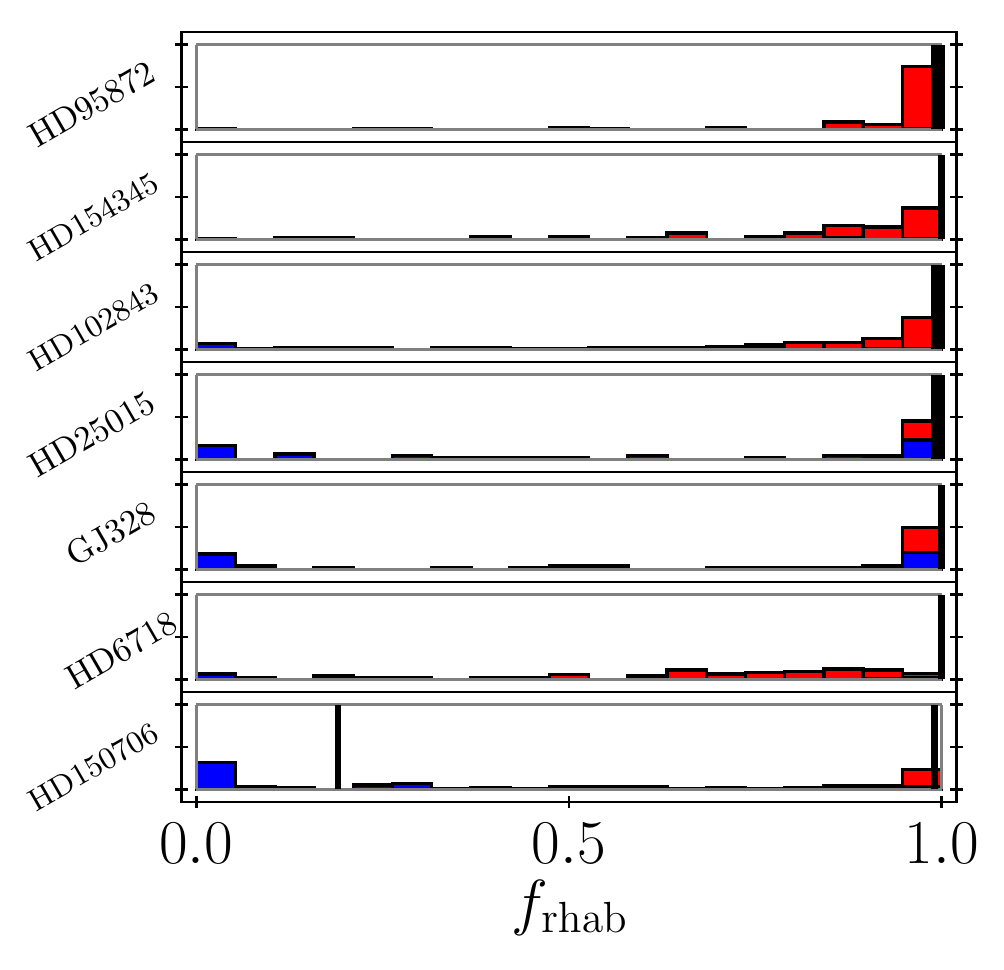}
\caption{The binned, resampled data used to produce the $f_{\rm rhab}$-distributions in figure~\ref{fig:frhab}. The histograms are normalized such that for each system the blue and the red bins together add up to one. The blue bins show the number of 3E runs and the red bins show the 4H runs. The black lines show the bounds of the $f_{\rm hab,1P}$-values.}
\label{fig:hists}
\end{figure}

We also highlight in Figure~\ref{fig:obs} some other systems that 
offer good prospects for follow up. HD~6718 and HD~150706 have 
large uncertainties on their eccentricities, and if the true 
eccentricities are at the lower end of the allowed ranges then 
their resilient habitability is high (see Figure~\ref{fig:3lines}). 
HD~25015 and GJ~328 are of high eccentricity ($e\sim0.4$) but distant 
from the HZ. They show strongly bimodal distributions of $f_\mathrm{rhab}$: 
if their past history involved ejection of an equal-mass planet then 
$f_\mathrm{rhab}\approx 0$, but if they have ejected a lower-mass planet 
then their resilient habitability is high. The $f_\mathrm{rhab}$ 
distribtions of these seven systems of interest 
are shown in Figure~\ref{fig:hists}.

In the introduction we made a point about utilizing $Gaia$ to constrain 
the orbital plane of the giant planets and using that to identify 
edge-on systems that will also show a strong RV signal from any 
potential Earth-like planet. The systems we have studied, typically 
at distances of a few 10s of pc, will have a good enough signal-to-noise 
to detect the astrometric reflex motion of the star from the known 
Jovian planets \citep[][figure~2]{Ranalli+18}. Furthermore, 
the orbital inclination of the giant planet in these systems will 
be constrained to $\sigma_{\cos I} \sim 0.1$ \citep[][figure~9 and table~5]{Ranalli+18}.

Finally, we consider how the orbit of the stars through the Galaxy may affect 
the habitability of the systems. \cite{Kokaia2019} showed how the passage of 
a star and its planetary system through giant molecular clouds -- which may 
trigger mass extictions -- depends on the star's orbit in the Galaxy. 
We obtain the positions and velocities of the 7 highlighted systems from 
Gaia DR2~\citep{GaiaDR2} and integrate them to determine their Galactic orbits
(see \citealt{Kokaia2019} for details). We find that for all of the systems, 
their orbits in the Galaxy lead to excursions of more than 100~pc from the 
Galactic plane. Comparing to~\citep{Kokaia2019} we find that all of them have 
encounters with giant molecular clouds at a similar rate to the Sun: 
between once every 500 Myr and 3 Gyr.

\section{Summary}\label{sec:su}

We have selected 34 systems containing a single massive planet orbiting an 
FGK-dwarf beyond the habitable zone (HZ). We simulated the 
dynamical history of these systems considering initial multiple-planet systems comprising
either three equal-mass planets (3E) or four planets having a mass hierarchy of the four gas giants
of our Solar System (4H).  These systems pass through a phase of instability where planets
are ejected via planet--planet scattering leaving other planets on more bound and more eccentric
orbits.  One can probe the entire observed eccentricity range with 3E and 4H systems. Low-eccentricity
planets are produced mostly by 4H systems, whereas all systems containing a planet with an
eccentricity, $ e > 0.5$ are derived exclusively from 3E systems.

We select the subset of the runs which give eccentricities of the inner planet matching observations.
Our simulations include test particles in the habitable zone, to represent 
terrestrial planets. We measure the fraction of test particles located within the HZ which survive the dynamical evolution,
and term this fraction the resilient habitability, $f_{\rm rhab}$. We also compute the fraction 
of test particles within the HZ which would be removed from the HZ by the single planet
as observed today, and refer to this as the present-day habitability, $f_{\rm hab,1P}$.

 For the 3E runs, the resulting instability tends to be much more violent resulting in higher eccentricities and planets often entering the HZ and clearing it out. For 4H, on the other hand, the two 
most massive planets rarely enter the HZ with the most massive one only doing so in systems where the observed planet is very close to the HZ boundary. This leaves the HZ mostly intact as the lower mass planets cannot directly eject the test particles.
The systems, shown in figure~\ref{fig:current_systems}, are placed into one of three groups:
\begin{itemize}
\item[] Group 1: Systems where the planet has a much larger semi-major axis than the outer 
  edge of the HZ and/or has a low eccentricity. These systems all have high values 
  of $f_{\rm hab,1P}$ and the median values of the $f_{\rm rhab}$ distribution also tend to be 
  high. However, some of the systems in the group have distributions of $f_\mathrm{rhab}$ 
  that go down to much lower values as can be seen in figure~\ref{fig:frhab}. 
  In all cases, $f_{\rm rhab}$ tends
  to be lower than $f_{\rm hab,1P}$. 
  In other words, {\it history matters}: the dynamical past is more 
  damaging than present perturbations of the observed planet.

\item[] Group 2: Systems where the planet has a slightly larger semi-major axis than the outer edge of the HZ and/or has a high eccentricity which brings it very close to the HZ. These planets are much more damaging to 
objects within the HZ. Systems in Group 2 generally show much lower values of $f_{\rm hab,1P}$
than seen in Group 1. Values of $f_{\rm rhab}$ are even lower owing to the extra damaging effects
of the earlier phase of planet--planet scattering.

\item[] Group 3: Systems that could fit in either Group 1 or Group 2, due to the planets having large uncertainties in their eccentricities which leads to a large range of possible $f_{\rm hab,1P}$ and $f_{\rm rhab}$ values. For these systems we recalculate the $f_{\rm rhab}$ distribution by considering three narrower ranges
of eccentricities centred around the mean and $+/-1\sigma$ values of the observed eccentricities
as shown in figure~\ref{fig:3lines}. We see two systems that show very large differences between 
the distribution of $f_\mathrm{rhab}$ if the eccentricity is high, and that if it is low; these are 
HD~6718 and HD~150706, and therefore better determination of these eccentricities is needed.
\end{itemize}

 Even in the 4H runs where the most-massive planet does not enter the HZ, the planet can do significant damage to the HZ when its semi-major axis changes during the scattering phase, as then  mean-motion resonances can sweep over the HZ and pump up the eccentricity of the test particles until they have a close encounter with the gas giant and get ejected or until the test particle hits the star. This is consistent with models for the depletion of the asteroid belt in the Solar System~\citep{Izidoro2016, Clement2019}.

In contrast to the mean-motion resonances, the secular resonances very rarely do any damage to the habitable zone. This is because for most systems, having the secular resonances inside the habitable zone requires the two planets to be 4-7 mutual hill radii apart. Whilst this is a stable configuration for circular coplanar orbits, it is rarely stable after an instability. The planets can in principle end up at the right semi-major axes but they will have eccentric orbits which generally leads to either one of them eventually getting ejected or a reconfiguration such that the secular resonances move away from the HZ.

We provide a list of systems containing at least one known giant planets most likely to be able to host an Earth-like planet:  HD~95872, HD~154345, HD~102843, HD~25015, GJ~328, HD~6718 and HD~150706. 
HD~95872 is the clearly best candidate as can be seen in figure~\ref{fig:obs}.

\section*{Acknowledgments}
The authors are supported by the project grant 2014.0017 ``IMPACT" from the Knut and Alice Wallenberg Foundation. The simulations were performed on resources provided by the Swedish National Infrastructure for Computing (SNIC) at Lunarc, which we can contribute thanks to grants from The Royal Physiographic Society of Lund. We thank Ross P.~Church for comments on the manuscript. We also thank the referee for their feedback which helped improve the paper.

\bibliography{Prefs}

\begin{thebibliography}{}
\makeatletter
\relax
\def\mn@urlcharsother{\let\do\@makeother \do\$\do\&\do\#\do\^\do\_\do\%\do\~}
\def\mn@doi{\begingroup\mn@urlcharsother \@ifnextchar [ {\mn@doi@}
  {\mn@doi@[]}}
\def\mn@doi@[#1]#2{\def\@tempa{#1}\ifx\@tempa\@empty \href
  {http://dx.doi.org/#2} {doi:#2}\else \href {http://dx.doi.org/#2} {#1}\fi
  \endgroup}
\def\mn@eprint#1#2{\mn@eprint@#1:#2::\@nil}
\def\mn@eprint@arXiv#1{\href {http://arxiv.org/abs/#1} {{\tt arXiv:#1}}}
\def\mn@eprint@dblp#1{\href {http://dblp.uni-trier.de/rec/bibtex/#1.xml}
  {dblp:#1}}
\def\mn@eprint@#1:#2:#3:#4\@nil{\def\@tempa {#1}\def\@tempb {#2}\def\@tempc
  {#3}\ifx \@tempc \@empty \let \@tempc \@tempb \let \@tempb \@tempa \fi \ifx
  \@tempb \@empty \def\@tempb {arXiv}\fi \@ifundefined
  {mn@eprint@\@tempb}{\@tempb:\@tempc}{\expandafter \expandafter \csname
  mn@eprint@\@tempb\endcsname \expandafter{\@tempc}}}

\bibitem[\protect\citeauthoryear{{Abramov} \& {Mojzsis}}{{Abramov} \&
  {Mojzsis}}{2009}]{Abramov2009}
{Abramov} O.,  {Mojzsis} S.~J.,  2009, \mn@doi [\nat] {10.1038/nature08015},
  \href {https://ui.adsabs.harvard.edu/abs/2009Natur.459..419A} {459, 419}

\bibitem[\protect\citeauthoryear{{Agnew}, {Maddison}, {Thilliez}  \&
  {Horner}}{{Agnew} et~al.}{2017}]{Agnew2017}
{Agnew} M.~T.,  {Maddison} S.~T.,  {Thilliez} E.,   {Horner} J.,  2017, \mn@doi
  [\mnras] {10.1093/mnras/stx1449}, \href
  {http://adsabs.harvard.edu/abs/2017MNRAS.471.4494A} {471, 4494}

\bibitem[\protect\citeauthoryear{{Agnew}, {Maddison}  \& {Horner}}{{Agnew}
  et~al.}{2018}]{Agnew2018}
{Agnew} M.~T.,  {Maddison} S.~T.,   {Horner} J.,  2018, \mn@doi [\mnras]
  {10.1093/mnras/sty868}, \href
  {http://adsabs.harvard.edu/abs/2018MNRAS.477.3646A} {477, 3646}

\bibitem[\protect\citeauthoryear{{Anglada-Escud{\'e}}, {L{\'o}pez-Morales}  \&
  {Chambers}}{{Anglada-Escud{\'e}} et~al.}{2010}]{Anglada2010}
{Anglada-Escud{\'e}} G.,  {L{\'o}pez-Morales} M.,   {Chambers} J.~E.,  2010,
  \mn@doi [\apj] {10.1088/0004-637X/709/1/168}, \href
  {https://ui.adsabs.harvard.edu/abs/2010ApJ...709..168A} {709, 168}

\bibitem[\protect\citeauthoryear{{Antoniadou} \& {Voyatzis}}{{Antoniadou} \&
  {Voyatzis}}{2016}]{Antoniadou2016}
{Antoniadou} K.~I.,  {Voyatzis} G.,  2016, \mn@doi [\mnras]
  {10.1093/mnras/stw1553}, \href
  {https://ui.adsabs.harvard.edu/abs/2016MNRAS.461.3822A} {461, 3822}

\bibitem[\protect\citeauthoryear{{Bashi}, {Helled}, {Zucker}  \&
  {Mordasini}}{{Bashi} et~al.}{2017}]{Bashi2017}
{Bashi} D.,  {Helled} R.,  {Zucker} S.,   {Mordasini} C.,  2017, \mn@doi [\aap]
  {10.1051/0004-6361/201629922}, \href
  {http://adsabs.harvard.edu/abs/2017A%26A...604A..83B} {604, A83}

\bibitem[\protect\citeauthoryear{{Batygin} \& {Laughlin}}{{Batygin} \&
  {Laughlin}}{2008}]{Batigyn2008}
{Batygin} K.,  {Laughlin} G.,  2008, \mn@doi [\apj] {10.1086/589232}, \href
  {https://ui.adsabs.harvard.edu/abs/2008ApJ...683.1207B} {683, 1207}

\bibitem[\protect\citeauthoryear{{Batygin} \& {Laughlin}}{{Batygin} \&
  {Laughlin}}{2015}]{Batygin2015}
{Batygin} K.,  {Laughlin} G.,  2015, \mn@doi [Proceedings of the National
  Academy of Science] {10.1073/pnas.1423252112}, \href
  {https://ui.adsabs.harvard.edu/abs/2015PNAS..112.4214B} {112, 4214}

\bibitem[\protect\citeauthoryear{{Benedict} et~al.,}{{Benedict}
  et~al.}{2006}]{Benedict2006}
{Benedict} G.~F.,  et~al., 2006, \mn@doi [\aj] {10.1086/508323}, \href
  {https://ui.adsabs.harvard.edu/abs/2006AJ....132.2206B} {132, 2206}

\bibitem[\protect\citeauthoryear{{Bitsch}, {Crida}, {Libert}  \&
  {Lega}}{{Bitsch} et~al.}{2013}]{Bitsch2013}
{Bitsch} B.,  {Crida} A.,  {Libert} A.~S.,   {Lega} E.,  2013, \mn@doi [\aap]
  {10.1051/0004-6361/201220310}, \href
  {https://ui.adsabs.harvard.edu/abs/2013A&A...555A.124B} {555, A124}

\bibitem[\protect\citeauthoryear{{Bitsch}, {Izidoro}, {Johansen}, {Raymond},
  {Morbidelli}, {Lambrechts}  \& {Jacobson}}{{Bitsch}
  et~al.}{2019}]{Bitsch2019}
{Bitsch} B.,  {Izidoro} A.,  {Johansen} A.,  {Raymond} S.~N.,  {Morbidelli} A.,
   {Lambrechts} M.,   {Jacobson} S.~A.,  2019, \mn@doi [\aap]
  {10.1051/0004-6361/201834489}, \href
  {https://ui.adsabs.harvard.edu/abs/2019A&A...623A..88B} {623, A88}

\bibitem[\protect\citeauthoryear{{Boisse} et~al.,}{{Boisse}
  et~al.}{2012}]{Boisse2012}
{Boisse} I.,  et~al., 2012, \mn@doi [\aap] {10.1051/0004-6361/201118419}, \href
  {http://adsabs.harvard.edu/abs/2012A%26A...545A..55B} {545, A55}

\bibitem[\protect\citeauthoryear{{Bolmont}, {Libert}, {Leconte}  \&
  {Selsis}}{{Bolmont} et~al.}{2016}]{Bolmont2016}
{Bolmont} E.,  {Libert} A.-S.,  {Leconte} J.,   {Selsis} F.,  2016, \mn@doi
  [\aap] {10.1051/0004-6361/201628073}, \href
  {https://ui.adsabs.harvard.edu/abs/2016A&A...591A.106B} {591, A106}

\bibitem[\protect\citeauthoryear{{Bryan}, {Knutson}, {Lee}, {Fulton},
  {Batygin}, {Ngo}  \& {Meshkat}}{{Bryan} et~al.}{2019}]{Bryan2019}
{Bryan} M.~L.,  {Knutson} H.~A.,  {Lee} E.~J.,  {Fulton} B.~J.,  {Batygin} K.,
  {Ngo} H.,   {Meshkat} T.,  2019, \mn@doi [\aj] {10.3847/1538-3881/aaf57f},
  \href {https://ui.adsabs.harvard.edu/abs/2019AJ....157...52B} {157, 52}

\bibitem[\protect\citeauthoryear{{Buchhave}, {Bitsch}, {Johansen}, {Latham},
  {Bizzarro}, {Bieryla}  \& {Kipping}}{{Buchhave} et~al.}{2018}]{Buchhave2018}
{Buchhave} L.~A.,  {Bitsch} B.,  {Johansen} A.,  {Latham} D.~W.,  {Bizzarro}
  M.,  {Bieryla} A.,   {Kipping} D.~M.,  2018, \mn@doi [\apj]
  {10.3847/1538-4357/aaafca}, \href
  {https://ui.adsabs.harvard.edu/abs/2018ApJ...856...37B} {856, 37}

\bibitem[\protect\citeauthoryear{{Carrera}, {Davies}  \& {Johansen}}{{Carrera}
  et~al.}{2016}]{Carrera2016}
{Carrera} D.,  {Davies} M.~B.,   {Johansen} A.,  2016, \mn@doi [\mnras]
  {10.1093/mnras/stw2218}, \href
  {http://adsabs.harvard.edu/abs/2016MNRAS.463.3226C} {463, 3226}

\bibitem[\protect\citeauthoryear{{Carter-Bond}, {O'Brien}  \&
  {Raymond}}{{Carter-Bond} et~al.}{2014}]{CarterBond2014}
{Carter-Bond} J.~C.,  {O'Brien} D.~P.,   {Raymond} S.~N.,  2014, in
  {Haghighipour} N.,  ed.,  IAU Symposium Vol. 293, Formation, Detection, and
  Characterization of Extrasolar Habitable Planets. pp 229--234,
  \mn@doi{10.1017/S174392131301288X}

\bibitem[\protect\citeauthoryear{{Chambers}}{{Chambers}}{1999}]{Chambers1999}
{Chambers} J.~E.,  1999, \mn@doi [\mnras] {10.1046/j.1365-8711.1999.02379.x},
  \href {http://adsabs.harvard.edu/abs/1999MNRAS.304..793C} {304, 793}

\bibitem[\protect\citeauthoryear{{Chambers}}{{Chambers}}{2001}]{Chambers2001}
{Chambers} J.~E.,  2001, \mn@doi [\icarus] {10.1006/icar.2001.6639}, \href
  {https://ui.adsabs.harvard.edu/abs/2001Icar..152..205C} {152, 205}

\bibitem[\protect\citeauthoryear{{Chambers}, {Wetherill}  \& {Boss}}{{Chambers}
  et~al.}{1996}]{Chambers1996}
{Chambers} J.~E.,  {Wetherill} G.~W.,   {Boss} A.~P.,  1996, \mn@doi [\icarus]
  {10.1006/icar.1996.0019}, \href
  {https://ui.adsabs.harvard.edu/abs/1996Icar..119..261C} {119, 261}

\bibitem[\protect\citeauthoryear{{Childs}, {Quintana}, {Barclay}  \&
  {Steffen}}{{Childs} et~al.}{2019}]{Childs2019}
{Childs} A.~C.,  {Quintana} E.,  {Barclay} T.,   {Steffen} J.~H.,  2019,
  \mn@doi [\mnras] {10.1093/mnras/stz385}, \href
  {https://ui.adsabs.harvard.edu/abs/2019MNRAS.485..541C} {485, 541}

\bibitem[\protect\citeauthoryear{{Chyba}}{{Chyba}}{1990}]{Chyba1990}
{Chyba} C.~F.,  1990, \mn@doi [\nat] {10.1038/343129a0}, \href
  {https://ui.adsabs.harvard.edu/abs/1990Natur.343..129C} {343, 129}

\bibitem[\protect\citeauthoryear{{Clement}, {Raymond}  \& {Kaib}}{{Clement}
  et~al.}{2019}]{Clement2019}
{Clement} M.~S.,  {Raymond} S.~N.,   {Kaib} N.~A.,  2019, \mn@doi [The
  Astronomical Journal] {10.3847/1538-3881/aaf21e}, \href
  {https://ui.adsabs.harvard.edu/abs/2019AJ....157...38C} {157, 38}

\bibitem[\protect\citeauthoryear{{Davies}, {Adams}, {Armitage}, {Chambers},
  {Ford}, {Morbidelli}, {Raymond}  \& {Veras}}{{Davies}
  et~al.}{2014}]{Davies2014}
{Davies} M.~B.,  {Adams} F.~C.,  {Armitage} P.,  {Chambers} J.,  {Ford} E.,
  {Morbidelli} A.,  {Raymond} S.~N.,   {Veras} D.,  2014, in {Beuther} H.,
  {Klessen} R.~S.,  {Dullemond} C.~P.,   {Henning} T.,  eds, Protostars and
  Planets VI. p.~787 (\mn@eprint {arXiv} {1311.6816}),
  \mn@doi{10.2458/azu_uapress_9780816531240-ch034}

\bibitem[\protect\citeauthoryear{{Denham}, {Naoz}, {Hoang}, {Stephan}  \&
  {Farr}}{{Denham} et~al.}{2019}]{Denham2019}
{Denham} P.,  {Naoz} S.,  {Hoang} B.-M.,  {Stephan} A.~P.,   {Farr} W.~M.,
  2019, \mn@doi [\mnras] {10.1093/mnras/sty2830}, \href
  {https://ui.adsabs.harvard.edu/abs/2019MNRAS.482.4146D} {482, 4146}

\bibitem[\protect\citeauthoryear{{Dressing}, {Spiegel}, {Scharf}, {Menou}  \&
  {Raymond}}{{Dressing} et~al.}{2010}]{Dressing2010}
{Dressing} C.~D.,  {Spiegel} D.~S.,  {Scharf} C.~A.,  {Menou} K.,   {Raymond}
  S.~N.,  2010, \mn@doi [\apj] {10.1088/0004-637X/721/2/1295}, \href
  {https://ui.adsabs.harvard.edu/abs/2010ApJ...721.1295D} {721, 1295}

\bibitem[\protect\citeauthoryear{{Endl} et~al.,}{{Endl}
  et~al.}{2016}]{Endl2016}
{Endl} M.,  et~al., 2016, \mn@doi [\apj] {10.3847/0004-637X/818/1/34}, \href
  {https://ui.adsabs.harvard.edu/abs/2016ApJ...818...34E} {818, 34}

\bibitem[\protect\citeauthoryear{{Feng} et~al.,}{{Feng}
  et~al.}{2019}]{Feng2019}
{Feng} F.,  et~al., 2019, \mn@doi [The Astrophysical Journal Supplement Series]
  {10.3847/1538-4365/ab1b16}, \href
  {https://ui.adsabs.harvard.edu/abs/2019ApJS..242...25F} {242, 25}

\bibitem[\protect\citeauthoryear{{Ford} \& {Rasio}}{{Ford} \&
  {Rasio}}{2008}]{Ford2008}
{Ford} E.~B.,  {Rasio} F.~A.,  2008, \mn@doi [\apj] {10.1086/590926}, \href
  {https://ui.adsabs.harvard.edu/abs/2008ApJ...686..621F} {686, 621}

\bibitem[\protect\citeauthoryear{{Gaia Collaboration} et~al.,}{{Gaia
  Collaboration} et~al.}{2018}]{GaiaDR2}
{Gaia Collaboration} et~al., 2018, \mn@doi [Astronomy and Astrophysics]
  {10.1051/0004-6361/201833051}, \href
  {https://ui.adsabs.harvard.edu/abs/2018A&A...616A...1G} {616, A1}

\bibitem[\protect\citeauthoryear{{Georgakarakos}, {Eggl}  \&
  {Dobbs-Dixon}}{{Georgakarakos} et~al.}{2018}]{Georgakarakos2018}
{Georgakarakos} N.,  {Eggl} S.,   {Dobbs-Dixon} I.,  2018, \mn@doi [\apj]
  {10.3847/1538-4357/aaaf72}, \href
  {https://ui.adsabs.harvard.edu/abs/2018ApJ...856..155G} {856, 155}

\bibitem[\protect\citeauthoryear{{Gilmozzi} \& {Spyromilio}}{{Gilmozzi} \&
  {Spyromilio}}{2007}]{Gilmozzi2007}
{Gilmozzi} R.,  {Spyromilio} J.,  2007, The Messenger, \href
  {https://ui.adsabs.harvard.edu/abs/2007Msngr.127...11G} {127, 11}

\bibitem[\protect\citeauthoryear{{Giuppone}, {Morais}  \& {Correia}}{{Giuppone}
  et~al.}{2013}]{Giuppone2013}
{Giuppone} C.~A.,  {Morais} M.~H.~M.,   {Correia} A.~C.~M.,  2013, \mn@doi
  [\mnras] {10.1093/mnras/stt1831}, \href
  {https://ui.adsabs.harvard.edu/abs/2013MNRAS.436.3547G} {436, 3547}

\bibitem[\protect\citeauthoryear{{Gladman}}{{Gladman}}{1993}]{Gladman1993}
{Gladman} B.,  1993, \mn@doi [\icarus] {10.1006/icar.1993.1169}, \href
  {http://adsabs.harvard.edu/abs/1993Icar..106..247G} {106, 247}

\bibitem[\protect\citeauthoryear{{Grazier}}{{Grazier}}{2016}]{Grazier2016}
{Grazier} K.~R.,  2016, \mn@doi [Astrobiology] {10.1089/ast.2015.1321}, \href
  {https://ui.adsabs.harvard.edu/abs/2016AsBio..16...23G} {16, 23}

\bibitem[\protect\citeauthoryear{{Hadden} \& {Lithwick}}{{Hadden} \&
  {Lithwick}}{2018}]{Hadden2018}
{Hadden} S.,  {Lithwick} Y.,  2018, \mn@doi [\aj] {10.3847/1538-3881/aad32c},
  \href {https://ui.adsabs.harvard.edu/abs/2018AJ....156...95H} {156, 95}

\bibitem[\protect\citeauthoryear{{Halliday}}{{Halliday}}{2008}]{Halliday2008}
{Halliday} A.~N.,  2008, \mn@doi [Philosophical Transactions of the Royal
  Society of London Series A] {10.1098/rsta.2008.0209}, \href
  {https://ui.adsabs.harvard.edu/abs/2008RSPTA.366.4163H} {366, 4163}

\bibitem[\protect\citeauthoryear{{Hansen}}{{Hansen}}{2009}]{Hansen2009}
{Hansen} B. M.~S.,  2009, \mn@doi [\apj] {10.1088/0004-637X/703/1/1131}, \href
  {https://ui.adsabs.harvard.edu/abs/2009ApJ...703.1131H} {703, 1131}

\bibitem[\protect\citeauthoryear{{Hansen}}{{Hansen}}{2017}]{Hansen2017}
{Hansen} B. M.~S.,  2017, \mn@doi [\mnras] {10.1093/mnras/stx182}, \href
  {https://ui.adsabs.harvard.edu/abs/2017MNRAS.467.1531H} {467, 1531}

\bibitem[\protect\citeauthoryear{{Hara}, {Bou{\'e}}, {Laskar}, {Delisle}  \&
  {Unger}}{{Hara} et~al.}{2019}]{Hara2019}
{Hara} N.~C.,  {Bou{\'e}} G.,  {Laskar} J.,  {Delisle} J.~B.,   {Unger} N.,
  2019, \mn@doi [\mnras] {10.1093/mnras/stz1849}, \href
  {https://ui.adsabs.harvard.edu/abs/2019MNRAS.489..738H} {489, 738}

\bibitem[\protect\citeauthoryear{{Horner} \& {Jones}}{{Horner} \&
  {Jones}}{2008}]{Horner2008}
{Horner} J.,  {Jones} B.~W.,  2008, \mn@doi [International Journal of
  Astrobiology] {10.1017/S1473550408004187}, \href
  {https://ui.adsabs.harvard.edu/abs/2008IJAsB...7..251H} {7, 251}

\bibitem[\protect\citeauthoryear{{Horner} \& {Jones}}{{Horner} \&
  {Jones}}{2009}]{Horner2009}
{Horner} J.,  {Jones} B.~W.,  2009, \mn@doi [International Journal of
  Astrobiology] {10.1017/S1473550408004357}, \href
  {http://adsabs.harvard.edu/abs/2009IJAsB...8...75H} {8, 75}

\bibitem[\protect\citeauthoryear{{Horner}, {Jones}  \& {Chambers}}{{Horner}
  et~al.}{2010}]{Horner2010}
{Horner} J.,  {Jones} B.~W.,   {Chambers} J.,  2010, \mn@doi [International
  Journal of Astrobiology] {10.1017/S1473550409990346}, \href
  {http://adsabs.harvard.edu/abs/2010IJAsB...9....1H} {9, 1}

\bibitem[\protect\citeauthoryear{{Innanen}, {Zheng}, {Mikkola}  \&
  {Valtonen}}{{Innanen} et~al.}{1997}]{Innanen1997}
{Innanen} K.~A.,  {Zheng} J.~Q.,  {Mikkola} S.,   {Valtonen} M.~J.,  1997,
  \mn@doi [\aj] {10.1086/118405}, \href
  {https://ui.adsabs.harvard.edu/abs/1997AJ....113.1915I} {113, 1915}

\bibitem[\protect\citeauthoryear{{Izidoro}, {Raymond}, {Pierens}, {Morbidelli},
  {Winter}  \& {Nesvorny`}}{{Izidoro} et~al.}{2016}]{Izidoro2016}
{Izidoro} A.,  {Raymond} S.~N.,  {Pierens} A.,  {Morbidelli} A.,  {Winter}
  O.~C.,   {Nesvorny`} D.,  2016, \mn@doi [The Astrophysical Journal]
  {10.3847/1538-4357/833/1/40}, \href
  {https://ui.adsabs.harvard.edu/abs/2016ApJ...833...40I} {833, 40}

\bibitem[\protect\citeauthoryear{{Jacobsen}}{{Jacobsen}}{2005}]{Jacobsen2005}
{Jacobsen} S.~B.,  2005, \mn@doi [Annual Review of Earth and Planetary
  Sciences] {10.1146/annurev.earth.33.092203.122614}, \href
  {https://ui.adsabs.harvard.edu/abs/2005AREPS..33..531J} {33, 531}

\bibitem[\protect\citeauthoryear{{Jenkins} et~al.,}{{Jenkins}
  et~al.}{2017}]{Jenkins2017}
{Jenkins} J.~S.,  et~al., 2017, \mn@doi [\mnras] {10.1093/mnras/stw2811}, \href
  {https://ui.adsabs.harvard.edu/abs/2017MNRAS.466..443J} {466, 443}

\bibitem[\protect\citeauthoryear{{Johansen}, {Davies}, {Church}  \&
  {Holmelin}}{{Johansen} et~al.}{2012}]{Johansen2012}
{Johansen} A.,  {Davies} M.~B.,  {Church} R.~P.,   {Holmelin} V.,  2012,
  \mn@doi [\apj] {10.1088/0004-637X/758/1/39}, \href
  {http://adsabs.harvard.edu/abs/2012ApJ...758...39J} {758, 39}

\bibitem[\protect\citeauthoryear{{Jones}, {Sleep}  \& {Chambers}}{{Jones}
  et~al.}{2001}]{Jones2001}
{Jones} B.~W.,  {Sleep} P.~N.,   {Chambers} J.~E.,  2001, \mn@doi [\aap]
  {10.1051/0004-6361:20000078}, \href
  {https://ui.adsabs.harvard.edu/abs/2001A&A...366..254J} {366, 254}

\bibitem[\protect\citeauthoryear{{Juri{\'c}} \& {Tremaine}}{{Juri{\'c}} \&
  {Tremaine}}{2008}]{Juric2008}
{Juri{\'c}} M.,  {Tremaine} S.,  2008, \mn@doi [\apj] {10.1086/590047}, \href
  {https://ui.adsabs.harvard.edu/abs/2008ApJ...686..603J} {686, 603}

\bibitem[\protect\citeauthoryear{{Kaib}, {Raymond}  \& {Duncan}}{{Kaib}
  et~al.}{2011}]{Kaib2011}
{Kaib} N.~A.,  {Raymond} S.~N.,   {Duncan} M.~J.,  2011, \mn@doi [\apjl]
  {10.1088/2041-8205/742/2/L24}, \href
  {https://ui.adsabs.harvard.edu/abs/2011ApJ...742L..24K} {742, L24}

\bibitem[\protect\citeauthoryear{{Kane} \& {Torres}}{{Kane} \&
  {Torres}}{2017}]{Kane2017}
{Kane} S.~R.,  {Torres} S.~M.,  2017, \mn@doi [\aj] {10.3847/1538-3881/aa8fce},
  \href {https://ui.adsabs.harvard.edu/abs/2017AJ....154..204K} {154, 204}

\bibitem[\protect\citeauthoryear{{Kokaia} \& {Davies}}{{Kokaia} \&
  {Davies}}{2019}]{Kokaia2019}
{Kokaia} G.,  {Davies} M.~B.,  2019, \mn@doi [Monthly Notices of the Royal
  Astronomical Society] {10.1093/mnras/stz813}, \href
  {https://ui.adsabs.harvard.edu/abs/2019MNRAS.tmp..945K} {p.~945}

\bibitem[\protect\citeauthoryear{{Kopparapu}, {Ramirez}, {SchottelKotte},
  {Kasting}, {Domagal-Goldman}  \& {Eymet}}{{Kopparapu}
  et~al.}{2014}]{Kopparapu2014}
{Kopparapu} R.~K.,  {Ramirez} R.~M.,  {SchottelKotte} J.,  {Kasting} J.~F.,
  {Domagal-Goldman} S.,   {Eymet} V.,  2014, \mn@doi [\apjl]
  {10.1088/2041-8205/787/2/L29}, \href
  {http://adsabs.harvard.edu/abs/2014ApJ...787L..29K} {787, L29}

\bibitem[\protect\citeauthoryear{{K{\"u}rster}, {Trifonov}, {Reffert},
  {Kostogryz}  \& {Rodler}}{{K{\"u}rster} et~al.}{2015}]{Kurster2015}
{K{\"u}rster} M.,  {Trifonov} T.,  {Reffert} S.,  {Kostogryz} N.~M.,   {Rodler}
  F.,  2015, \mn@doi [\aap] {10.1051/0004-6361/201525872}, \href
  {https://ui.adsabs.harvard.edu/abs/2015A&A...577A.103K} {577, A103}

\bibitem[\protect\citeauthoryear{{Laakso}, {Rantala}  \&
  {Kaasalainen}}{{Laakso} et~al.}{2006}]{Laakso2006}
{Laakso} T.,  {Rantala} J.,   {Kaasalainen} M.,  2006, \mn@doi [\aap]
  {10.1051/0004-6361:20065121}, \href
  {http://adsabs.harvard.edu/abs/2006A%26A...456..373L} {456, 373}

\bibitem[\protect\citeauthoryear{{Lambrechts}, {Morbidelli}, {Jacobson},
  {Johansen}, {Bitsch}, {Izidoro}  \& {Raymond}}{{Lambrechts}
  et~al.}{2019}]{Lambrechts2019}
{Lambrechts} M.,  {Morbidelli} A.,  {Jacobson} S.~A.,  {Johansen} A.,  {Bitsch}
  B.,  {Izidoro} A.,   {Raymond} S.~N.,  2019, \mn@doi [\aap]
  {10.1051/0004-6361/201834229}, \href
  {https://ui.adsabs.harvard.edu/abs/2019A&A...627A..83L} {627, A83}

\bibitem[\protect\citeauthoryear{{Laskar} \& {Gastineau}}{{Laskar} \&
  {Gastineau}}{2009}]{Laskar2009}
{Laskar} J.,  {Gastineau} M.,  2009, \mn@doi [\nat] {10.1038/nature08096},
  \href {https://ui.adsabs.harvard.edu/abs/2009Natur.459..817L} {459, 817}

\bibitem[\protect\citeauthoryear{{Leite}, {Martins}, {Molaro}, {Monai},
  {Alves}, {Silva}  \& {the ESPRESSO Science Team}}{{Leite}
  et~al.}{2018}]{Leite2018}
{Leite} A.~C.~O.,  {Martins} C.~J.~A.~P.,  {Molaro} P.,  {Monai} S.,  {Alves}
  C.~S.,  {Silva} T.~A.,   {the ESPRESSO Science Team} 2018, arXiv e-prints,
  \href {https://ui.adsabs.harvard.edu/abs/2018arXiv181206796L} {p.
  arXiv:1812.06796}

\bibitem[\protect\citeauthoryear{{Linsenmeier}, {Pascale}  \&
  {Lucarini}}{{Linsenmeier} et~al.}{2015}]{Linsenmeier2015}
{Linsenmeier} M.,  {Pascale} S.,   {Lucarini} V.,  2015, \mn@doi [\planss]
  {10.1016/j.pss.2014.11.003}, \href
  {https://ui.adsabs.harvard.edu/abs/2015P&SS..105...43L} {105, 43}

\bibitem[\protect\citeauthoryear{{Lissauer} et~al.,}{{Lissauer}
  et~al.}{2011}]{Lissauer2011}
{Lissauer} J.~J.,  et~al., 2011, \mn@doi [\apjs] {10.1088/0067-0049/197/1/8},
  \href {http://adsabs.harvard.edu/abs/2011ApJS..197....8L} {197, 8}

\bibitem[\protect\citeauthoryear{{Lucy} \& {Sweeney}}{{Lucy} \&
  {Sweeney}}{1971}]{LucySweeney71}
{Lucy} L.~B.,  {Sweeney} M.~A.,  1971, \mn@doi [\aj] {10.1086/111159}, \href
  {https://ui.adsabs.harvard.edu/abs/1971AJ.....76..544L} {76, 544}

\bibitem[\protect\citeauthoryear{{Machida}, {Kokubo}, {Inutsuka}  \&
  {Matsumoto}}{{Machida} et~al.}{2010}]{Machida2010}
{Machida} M.~N.,  {Kokubo} E.,  {Inutsuka} S.-I.,   {Matsumoto} T.,  2010,
  \mn@doi [\mnras] {10.1111/j.1365-2966.2010.16527.x}, \href
  {https://ui.adsabs.harvard.edu/abs/2010MNRAS.405.1227M} {405, 1227}

\bibitem[\protect\citeauthoryear{{Malmberg}, {Davies}  \& {Heggie}}{{Malmberg}
  et~al.}{2011}]{Malmberg2011}
{Malmberg} D.,  {Davies} M.~B.,   {Heggie} D.~C.,  2011, \mn@doi [\mnras]
  {10.1111/j.1365-2966.2010.17730.x}, \href
  {https://ui.adsabs.harvard.edu/abs/2011MNRAS.411..859M} {411, 859}

\bibitem[\protect\citeauthoryear{{Mardling} \& {Aarseth}}{{Mardling} \&
  {Aarseth}}{2001}]{MardlingAarseth01}
{Mardling} R.~A.,  {Aarseth} S.~J.,  2001, \mn@doi [\mnras]
  {10.1046/j.1365-8711.2001.03974.x}, \href
  {https://ui.adsabs.harvard.edu/abs/2001MNRAS.321..398M} {321, 398}

\bibitem[\protect\citeauthoryear{{Marmier} et~al.,}{{Marmier}
  et~al.}{2013}]{Marmier2013}
{Marmier} M.,  et~al., 2013, \mn@doi [\aap] {10.1051/0004-6361/201219639},
  \href {http://adsabs.harvard.edu/abs/2013A%26A...551A..90M} {551, A90}

\bibitem[\protect\citeauthoryear{{Matsumura}, {Ida}  \& {Nagasawa}}{{Matsumura}
  et~al.}{2013}]{Matsumura2013}
{Matsumura} S.,  {Ida} S.,   {Nagasawa} M.,  2013, \mn@doi [\apj]
  {10.1088/0004-637X/767/2/129}, \href
  {https://ui.adsabs.harvard.edu/abs/2013ApJ...767..129M} {767, 129}

\bibitem[\protect\citeauthoryear{{Menou} \& {Tabachnik}}{{Menou} \&
  {Tabachnik}}{2003}]{Menou2003}
{Menou} K.,  {Tabachnik} S.,  2003, \mn@doi [\apj] {10.1086/345359}, \href
  {https://ui.adsabs.harvard.edu/abs/2003ApJ...583..473M} {583, 473}

\bibitem[\protect\citeauthoryear{{Ment}, {Fischer}, {Bakos}, {Howard}  \&
  {Isaacson}}{{Ment} et~al.}{2018}]{Ment2018}
{Ment} K.,  {Fischer} D.~A.,  {Bakos} G.,  {Howard} A.~W.,   {Isaacson} H.,
  2018, \mn@doi [\aj] {10.3847/1538-3881/aae1f5}, \href
  {http://adsabs.harvard.edu/abs/2018AJ....156..213M} {156, 213}

\bibitem[\protect\citeauthoryear{{Morbidelli}, {Chambers}, {Lunine}, {Petit},
  {Robert}, {Valsecchi}  \& {Cyr}}{{Morbidelli} et~al.}{2000}]{Morbidelli2000}
{Morbidelli} A.,  {Chambers} J.,  {Lunine} J.~I.,  {Petit} J.~M.,  {Robert} F.,
   {Valsecchi} G.~B.,   {Cyr} K.~E.,  2000, \mn@doi [Meteoritics and Planetary
  Science] {10.1111/j.1945-5100.2000.tb01518.x}, \href
  {https://ui.adsabs.harvard.edu/abs/2000M&PS...35.1309M} {35, 1309}

\bibitem[\protect\citeauthoryear{{Mordasini}, {Alibert}  \& {Benz}}{{Mordasini}
  et~al.}{2009}]{Mordasini2015}
{Mordasini} C.,  {Alibert} Y.,   {Benz} W.,  2009, \mn@doi [\aap]
  {10.1051/0004-6361/200810301}, \href
  {https://ui.adsabs.harvard.edu/abs/2009A&A...501.1139M} {501, 1139}

\bibitem[\protect\citeauthoryear{{Moutou} et~al.,}{{Moutou}
  et~al.}{2011}]{Moutou2011}
{Moutou} C.,  et~al., 2011, \mn@doi [\aap] {10.1051/0004-6361/201015371}, \href
  {http://adsabs.harvard.edu/abs/2011A%26A...527A..63M} {527, A63}

\bibitem[\protect\citeauthoryear{{Mullally} et~al.,}{{Mullally}
  et~al.}{2015}]{Mullally2015}
{Mullally} F.,  et~al., 2015, \mn@doi [\apjs] {10.1088/0067-0049/217/2/31},
  \href {https://ui.adsabs.harvard.edu/abs/2015ApJS..217...31M} {217, 31}

\bibitem[\protect\citeauthoryear{{Murray} \& {Dermott}}{{Murray} \&
  {Dermott}}{1999}]{Murray1999}
{Murray} C.~D.,  {Dermott} S.~F.,  1999, {Solar system dynamics}

\bibitem[\protect\citeauthoryear{{Mustill} \& {Wyatt}}{{Mustill} \&
  {Wyatt}}{2012}]{MustillWyatt12}
{Mustill} A.~J.,  {Wyatt} M.~C.,  2012, \mn@doi [\mnras]
  {10.1111/j.1365-2966.2011.19948.x}, \href
  {https://ui.adsabs.harvard.edu/abs/2012MNRAS.419.3074M} {419, 3074}

\bibitem[\protect\citeauthoryear{{Mustill}, {Davies}  \& {Johansen}}{{Mustill}
  et~al.}{2017}]{Mustill2017}
{Mustill} A.~J.,  {Davies} M.~B.,   {Johansen} A.,  2017, \mn@doi [\mnras]
  {10.1093/mnras/stx693}, \href
  {http://adsabs.harvard.edu/abs/2017MNRAS.468.3000M} {468, 3000}

\bibitem[\protect\citeauthoryear{{Mustill}, {Davies}  \& {Johansen}}{{Mustill}
  et~al.}{2018}]{Mustill2018}
{Mustill} A.~J.,  {Davies} M.~B.,   {Johansen} A.,  2018, \mn@doi [\mnras]
  {10.1093/mnras/sty1273}, \href
  {https://ui.adsabs.harvard.edu/abs/2018MNRAS.478.2896M} {478, 2896}

\bibitem[\protect\citeauthoryear{{O'Brien}, {Morbidelli}  \&
  {Levison}}{{O'Brien} et~al.}{2006}]{OBrien2006}
{O'Brien} D.~P.,  {Morbidelli} A.,   {Levison} H.~F.,  2006, \mn@doi [\icarus]
  {10.1016/j.icarus.2006.04.005}, \href
  {https://ui.adsabs.harvard.edu/abs/2006Icar..184...39O} {184, 39}

\bibitem[\protect\citeauthoryear{{O'Brien}, {Izidoro}, {Jacobson}, {Raymond}
  \& {Rubie}}{{O'Brien} et~al.}{2018}]{OBrien2018}
{O'Brien} D.~P.,  {Izidoro} A.,  {Jacobson} S.~A.,  {Raymond} S.~N.,   {Rubie}
  D.~C.,  2018, \mn@doi [Space Science Reviews] {10.1007/s11214-018-0475-8},
  \href {https://ui.adsabs.harvard.edu/abs/2018SSRv..214...47O} {214, 47}

\bibitem[\protect\citeauthoryear{{Pepe} et~al.,}{{Pepe}
  et~al.}{2010}]{Pepe2010}
{Pepe} F.~A.,  et~al., 2010, in Ground-based and Airborne Instrumentation for
  Astronomy III. p. 77350F, \mn@doi{10.1117/12.857122}

\bibitem[\protect\citeauthoryear{{Perryman}, {Hartman}, {Bakos}  \&
  {Lindegren}}{{Perryman} et~al.}{2014}]{Perryman2014}
{Perryman} M.,  {Hartman} J.,  {Bakos} G.~{\'A}.,   {Lindegren} L.,  2014,
  \mn@doi [\apj] {10.1088/0004-637X/797/1/14}, \href
  {https://ui.adsabs.harvard.edu/abs/2014ApJ...797...14P} {797, 14}

\bibitem[\protect\citeauthoryear{{Petrovich}}{{Petrovich}}{2015}]{Petrovich15}
{Petrovich} C.,  2015, \mn@doi [\apj] {10.1088/0004-637X/808/2/120}, \href
  {https://ui.adsabs.harvard.edu/abs/2015ApJ...808..120P} {808, 120}

\bibitem[\protect\citeauthoryear{{Piso} \& {Youdin}}{{Piso} \&
  {Youdin}}{2014}]{Piso2014}
{Piso} A.-M.~A.,  {Youdin} A.~N.,  2014, \mn@doi [\apj]
  {10.1088/0004-637X/786/1/21}, \href
  {https://ui.adsabs.harvard.edu/abs/2014ApJ...786...21P} {786, 21}

\bibitem[\protect\citeauthoryear{{Piso}, {Youdin}  \& {Murray-Clay}}{{Piso}
  et~al.}{2015}]{Piso2015}
{Piso} A.-M.~A.,  {Youdin} A.~N.,   {Murray-Clay} R.~A.,  2015, \mn@doi [\apj]
  {10.1088/0004-637X/800/2/82}, \href
  {https://ui.adsabs.harvard.edu/abs/2015ApJ...800...82P} {800, 82}

\bibitem[\protect\citeauthoryear{{Ragusa}, {Rosotti}, {Teyssandier}, {Booth},
  {Clarke}  \& {Lodato}}{{Ragusa} et~al.}{2018}]{Ragusa2018}
{Ragusa} E.,  {Rosotti} G.,  {Teyssandier} J.,  {Booth} R.,  {Clarke} C.~J.,
  {Lodato} G.,  2018, \mn@doi [\mnras] {10.1093/mnras/stx3094}, \href
  {https://ui.adsabs.harvard.edu/abs/2018MNRAS.474.4460R} {474, 4460}

\bibitem[\protect\citeauthoryear{{Ranalli}, {Hobbs}  \& {Lindegren}}{{Ranalli}
  et~al.}{2018}]{Ranalli+18}
{Ranalli} P.,  {Hobbs} D.,   {Lindegren} L.,  2018, \mn@doi [\aap]
  {10.1051/0004-6361/201730921}, \href
  {https://ui.adsabs.harvard.edu/abs/2018A&A...614A..30R} {614, A30}

\bibitem[\protect\citeauthoryear{{Rauer} et~al.,}{{Rauer}
  et~al.}{2014}]{Rauer2014}
{Rauer} H.,  et~al., 2014, \mn@doi [Experimental Astronomy]
  {10.1007/s10686-014-9383-4}, \href
  {https://ui.adsabs.harvard.edu/abs/2014ExA....38..249R} {38, 249}

\bibitem[\protect\citeauthoryear{{Raymond} \& {Izidoro}}{{Raymond} \&
  {Izidoro}}{2017}]{Raymond2017}
{Raymond} S.~N.,  {Izidoro} A.,  2017, \mn@doi [\icarus]
  {10.1016/j.icarus.2017.06.030}, \href
  {https://ui.adsabs.harvard.edu/abs/2017Icar..297..134R} {297, 134}

\bibitem[\protect\citeauthoryear{{Raymond}, {Mandell}  \&
  {Sigurdsson}}{{Raymond} et~al.}{2006}]{Raymond2006}
{Raymond} S.~N.,  {Mandell} A.~M.,   {Sigurdsson} S.,  2006, \mn@doi [Science]
  {10.1126/science.1130461}, \href
  {https://ui.adsabs.harvard.edu/abs/2006Sci...313.1413R} {313, 1413}

\bibitem[\protect\citeauthoryear{{Raymond}, {O'Brien}, {Morbidelli}  \&
  {Kaib}}{{Raymond} et~al.}{2009}]{Raymond2009}
{Raymond} S.~N.,  {O'Brien} D.~P.,  {Morbidelli} A.,   {Kaib} N.~A.,  2009,
  \mn@doi [\icarus] {10.1016/j.icarus.2009.05.016}, \href
  {https://ui.adsabs.harvard.edu/abs/2009Icar..203..644R} {203, 644}

\bibitem[\protect\citeauthoryear{{Rey} et~al.,}{{Rey} et~al.}{2017}]{Rey2017}
{Rey} J.,  et~al., 2017, \mn@doi [\aap] {10.1051/0004-6361/201630089}, \href
  {http://adsabs.harvard.edu/abs/2017A%26A...601A...9R} {601, A9}

\bibitem[\protect\citeauthoryear{{Rickman} et~al.,}{{Rickman}
  et~al.}{2019}]{Rickman2019}
{Rickman} E.~L.,  et~al., 2019, arXiv e-prints, \href
  {http://adsabs.harvard.edu/abs/2019arXiv190401573R} {}

\bibitem[\protect\citeauthoryear{{Robertson}, {Endl}, {Cochran}, {MacQueen}  \&
  {Boss}}{{Robertson} et~al.}{2013}]{Robertson2013}
{Robertson} P.,  {Endl} M.,  {Cochran} W.~D.,  {MacQueen} P.~J.,   {Boss}
  A.~P.,  2013, \mn@doi [\apj] {10.1088/0004-637X/774/2/147}, \href
  {https://ui.adsabs.harvard.edu/abs/2013ApJ...774..147R} {774, 147}

\bibitem[\protect\citeauthoryear{{Rowan} et~al.,}{{Rowan}
  et~al.}{2016}]{Rowan2016}
{Rowan} D.,  et~al., 2016, \mn@doi [\apj] {10.3847/0004-637X/817/2/104}, \href
  {http://adsabs.harvard.edu/abs/2016ApJ...817..104R} {817, 104}

\bibitem[\protect\citeauthoryear{{Shen} \& {Turner}}{{Shen} \&
  {Turner}}{2008}]{ShenTurner08}
{Shen} Y.,  {Turner} E.~L.,  2008, \mn@doi [\apj] {10.1086/590548}, \href
  {https://ui.adsabs.harvard.edu/abs/2008ApJ...685..553S} {685, 553}

\bibitem[\protect\citeauthoryear{{Shikita}, {Koyama}  \& {Yamada}}{{Shikita}
  et~al.}{2010}]{Shikita2010}
{Shikita} B.,  {Koyama} H.,   {Yamada} S.,  2010, \mn@doi [\apj]
  {10.1088/0004-637X/712/2/819}, \href
  {https://ui.adsabs.harvard.edu/abs/2010ApJ...712..819S} {712, 819}

\bibitem[\protect\citeauthoryear{{Skidmore}, {TMT International Science
  Development Teams}  \& {Science Advisory Committee}}{{Skidmore}
  et~al.}{2015}]{Skidmore2015}
{Skidmore} W.,  {TMT International Science Development Teams}  {Science
  Advisory Committee} T.,  2015, \mn@doi [Research in Astronomy and
  Astrophysics] {10.1088/1674-4527/15/12/001}, \href
  {https://ui.adsabs.harvard.edu/abs/2015RAA....15.1945S} {15, 1945}

\bibitem[\protect\citeauthoryear{{Sozzetti}, {Casertano}, {Lattanzi}, {Spagna},
  {Morbidelli}, {Pannunzio}, {Pourbaix}  \& {Queloz}}{{Sozzetti}
  et~al.}{2008}]{Sozzetti2008}
{Sozzetti} A.,  {Casertano} S.,  {Lattanzi} M.~G.,  {Spagna} A.,  {Morbidelli}
  R.,  {Pannunzio} R.,  {Pourbaix} D.,   {Queloz} D.,  2008, in {Jin} W.~J.,
  {Platais} I.,   {Perryman} M.~A.~C.,  eds,  IAU Symposium Vol. 248, A Giant
  Step: from Milli- to Micro-arcsecond Astrometry. pp 256--259 (\mn@eprint
  {arXiv} {0711.4903}), \mn@doi{10.1017/S1743921308019200}

\bibitem[\protect\citeauthoryear{{Spiegel}, {Raymond}, {Dressing}, {Scharf}  \&
  {Mitchell}}{{Spiegel} et~al.}{2010}]{Spiegel2010}
{Spiegel} D.~S.,  {Raymond} S.~N.,  {Dressing} C.~D.,  {Scharf} C.~A.,
  {Mitchell} J.~L.,  2010, \mn@doi [\apj] {10.1088/0004-637X/721/2/1308}, \href
  {http://adsabs.harvard.edu/abs/2010ApJ...721.1308S} {721, 1308}

\bibitem[\protect\citeauthoryear{{Stassun}, {Collins}  \& {Gaudi}}{{Stassun}
  et~al.}{2017}]{Stassun2017}
{Stassun} K.~G.,  {Collins} K.~A.,   {Gaudi} B.~S.,  2017, \mn@doi [\aj]
  {10.3847/1538-3881/aa5df3}, \href
  {http://adsabs.harvard.edu/abs/2017AJ....153..136S} {153, 136}

\bibitem[\protect\citeauthoryear{{Thommes}, {Matsumura}  \& {Rasio}}{{Thommes}
  et~al.}{2008}]{Thommes2008}
{Thommes} E.~W.,  {Matsumura} S.,   {Rasio} F.~A.,  2008, \mn@doi [Science]
  {10.1126/science.1159723}, \href
  {https://ui.adsabs.harvard.edu/abs/2008Sci...321..814T} {321, 814}

\bibitem[\protect\citeauthoryear{{Veras} \& {Armitage}}{{Veras} \&
  {Armitage}}{2005}]{VerasArmitage05}
{Veras} D.,  {Armitage} P.~J.,  2005, \mn@doi [\apjl] {10.1086/428831}, \href
  {https://ui.adsabs.harvard.edu/abs/2005ApJ...620L.111V} {620, L111}

\bibitem[\protect\citeauthoryear{{Ward}, {Brownlee}  \& {Krauss}}{{Ward}
  et~al.}{2000}]{Ward2000}
{Ward} P.~D.,  {Brownlee} D.,   {Krauss} L.,  2000, \mn@doi [Physics Today]
  {10.1063/1.1325239}, \href
  {https://ui.adsabs.harvard.edu/abs/2000PhT....53i..62W} {53, 62}

\bibitem[\protect\citeauthoryear{{Wittenmyer} et~al.,}{{Wittenmyer}
  et~al.}{2014}]{Wittenmyer2014}
{Wittenmyer} R.~A.,  et~al., 2014, \mn@doi [\apj]
  {10.1088/0004-637X/783/2/103}, \href
  {http://adsabs.harvard.edu/abs/2014ApJ...783..103W} {783, 103}

\bibitem[\protect\citeauthoryear{{Wittenmyer}, {Bergmann}, {Horner}, {Clark}
  \& {Kane}}{{Wittenmyer} et~al.}{2019}]{Wittenmyer2019}
{Wittenmyer} R.~A.,  {Bergmann} C.,  {Horner} J.,  {Clark} J.,   {Kane} S.~R.,
  2019, \mn@doi [\mnras] {10.1093/mnras/stz236}, \href
  {https://ui.adsabs.harvard.edu/abs/2019MNRAS.484.4230W} {484, 4230}

\bibitem[\protect\citeauthoryear{{Zakamska}, {Pan}  \& {Ford}}{{Zakamska}
  et~al.}{2011}]{Zakamska2011}
{Zakamska} N.~L.,  {Pan} M.,   {Ford} E.~B.,  2011, \mn@doi [\mnras]
  {10.1111/j.1365-2966.2010.17570.x}, \href
  {https://ui.adsabs.harvard.edu/abs/2011MNRAS.410.1895Z} {410, 1895}

\bibitem[\protect\citeauthoryear{{Zeebe}}{{Zeebe}}{2015}]{Zeebe2015}
{Zeebe} R.~E.,  2015, \mn@doi [\apj] {10.1088/0004-637X/798/1/8}, \href
  {https://ui.adsabs.harvard.edu/abs/2015ApJ...798....8Z} {798, 8}

\bibitem[\protect\citeauthoryear{{Zhu} \& {Wu}}{{Zhu} \& {Wu}}{2018}]{Zhu2018}
{Zhu} W.,  {Wu} Y.,  2018, \mn@doi [\aj] {10.3847/1538-3881/aad22a}, \href
  {http://adsabs.harvard.edu/abs/2018AJ....156...92Z} {156, 92}

\makeatother
\end{thebibliography}

\bibliographystyle{mnras}

\label{lastpage}
\end{document}